# COMPLETENESS CLASSES IN ALGEBRAIC COMPLEXITY THEORY

## PETER BÜRGISSER

ABSTRACT. The purpose of this overview is to explain the enormous impact of Les Valiant's eponymous short conference contribution from 1979 on the development of algebraic complexity.

## CONTENTS

*Date*: June 11, 2024.

2000 *Mathematics Subject Classification*. 68Q15, 68Q17.

The author was supported by the ERC under the European's Horizon 2020 research and innovation programme (grant agreement no. 787840). To appear in "Foundations of Computation and Machine Learning: The Work of Leslie Valiant", Ed. Rocco Servedio, Turing Award series, ACM.

## 1. Introduction

The short conference paper "Completeness classes in algebra" [Val79a], published by Leslie Valiant in 1979 and reproduced in this volume, had a profound effect on the further development of algebraic complexity theory. Its contents are of a similar flavour as Stephen Cook's seminal paper [Coo71]; interestingly, both papers never made it to journal version. Cook had introduced the complexity class NP and identified the satisfiability problem as a universal problem: it is NP-complete. The paramount importance of this achievement for all of computer science is well known and does not need to be discussed here.

In his seminal work [Val79a], Valiant introduced algebraic analogues VP and VNP of the complexity classes P and NP and proved that the evaluation of the permanent polynomials is VNP-complete. The proof relies on ingenious algorithmic reductions from combinatorial to algebraic problems. The computational problems studied in this framework are the evaluation of polynomial functions over some fixed field $\mathbb{F}$; the model of computation is the one of arithmetic circuits (or straight line programs) with the arithmetic operations as atomic operations. Valiant conjectured that the classes VP and VNP are different, which can be seen as a version of Cook's P $\neq$ NP conjecture, tailored to the framework of computations of polynomials. The analogy becomes even more apparent when thinking of the problems in P and NP as being given by families of Boolean functions, with the model of computation provided by (uniform) families of Boolean circuits. Replacing the basic Boolean operations or, and, not by the basic arithmetic operations addition, multiplication, negation, one gets polynomials instead of Boolean functions, and arithmetic circuits instead of Boolean circuits. Requiring that the size of the circuits and the degree of the polynomials computed grow at most polynomially, one naturally obtains the classes VP and VNP[1], defined with respect to a base field $\mathbb{F}$.

It is easy to show that VP = VNP over $\mathbb{F}_2$ implies P/poly = NP/poly, where the latter denotes the nonuniform version of P = NP, see [Val92]. The reason is that computations with polynomials can be efficiently simulated by Boolean circuits. There is a similar implication over general base fields $\mathbb{F}$, which relies on a sophisticated argument; see [Bür00b]. Therefore, if we want to prove P $\neq$ NP nonuniformly, we need to show VP $\neq$ VNP first. This observation is a compelling motivation to study the algebraic setting!

One of Valiant's major insights was that VP $\neq$ VNP can be rephrased as a quantitative question of an algebraic flavour, concerned with a comparison of the permanent with the determinant polynomials, which has no algorithmic contents whatsover: we describe this in detail in Section 2.5. While this may sound disappointing for algorithm lovers, it is good news for those who try to employ the highly developed arsenal of algebraic geometry to attack this problem. In fact, studying polynomials is a tradition in mathematics, which can look back on several hundred years. Over an algebraically closed base field (like the complex numbers $\mathbb{C}$) the situation is understood best. Algebraic geometry, widely considered to be the most sophisticated part of mathematics, exactly deals with systems of polynomials and their sets of solutions. Volker Strassen was the first to point out that major problems in (algebraic) complexity could be solved in the framework of algebraic geometry [Str84, Str87b]. The past twenty years have seen serious attempts in this direction (see Landsberg's book [Lan17], we will report about this in Section 7.

---

[1] It is common not to assume uniformity conditions on the circuits. The restriction on the growth of degree is essential.

The conference paper "Completeness classes in algebra" [Val79a] is closely connected to and appeared almost at the same time as two other of Leslie Valiant's ground breaking works, in which counting complexity theory was born. In the classic work [Val79b], Valiant had shown that the computation of the permanent of a matrix with 0,1 entries is a complete problem for the class #P of counting problems associated with nondeterministic polynomial time computation. This was surprising, since the corresponding decision problem is the perfect matching problem, which is well known to be solvable in polynomial time. Valiant defined the class #P in [Val79b] and identified a list of interesting #P-complete problems in [Val79c]. So while the articles [Val79b, Val79c] relied on the model of computation formalized by Turing machines, the related conference paper [Val79a] studied computations of polynomials over a field $\mathbb{F}$, modeling them by arithmetic formulas. The main results were the completeness of the permanent polynomials in a certain algebraically defined complexity class, nowadays called VNP, as well as the proof of completeness of the determinant polynomials in a smaller class thought to capture what is feasibly computable in this framework. In the subsequent work [Val82], Valiant further showed that the definition of VNP is robust with regard to various modifications: for instance, the more general model of arithmetic circuits instead of formulas leads to the same class. It should be noted that the definition of the complexity classes actually depends on the choice of the base field $\mathbb{F}$. The case of characteristic two (e.g., $\mathbb{F} = \mathbb{F}_2$) is exceptional since permanent and determinant coincide. A remarkable feature of [Val79a] is that the completeness results are proven for the simplest type of reduction one can think of: substitution! It turns out that this finding also applies to Boolean functions, which led to the joint publication [SV85] by Skyum and Valiant.

Valiant's short conference paper [Val79a] changed the course of algebraic complexity theory, but it took some time for the community to realize the significance of his contribution. Valiant's ideas were made more accessible in the surveys by von zur Gathen [Gat87] and Strassen [Str90]. Moreover, the very last chapter of the monograph [BCS97] by Bürgisser, Clausen and Shokrollahi contains an account of this theory, which was further elaborated in Bürgisser's book [Bür00a].

Traditionally, algebraic complexity focused on proving concrete complexity lower bounds for various problems related to the evaluation of polynomials and rational functions. Prominent examples include the discrete Fourier transform, matrix multiplication, and solving a system of linear equations; see [BCS97]. A major success was achieved by Strassen [Str73a], who proved an optimal nonlinear complexity lower bound, by relying on Bézout's theorem, a classical result of algebraic geometry. This was the first time ever a nonlinear complexity lower bound was shown. Combining this with an important reduction related to the computation of partial derivatives [BS83], more nonlinear complexity bounds were obtained. Unfortunately, despite intense efforts, these so far remained the only known nonlinear lower complexity bounds in unrestricted algebraic models of computation, see the recent book [Wig19, Chap. 12]. For instance, nobody knows how to prove a lower bound of order $n^2 \log n$ for the evaluation of the $n \times n$ determinant or $n \times n$ permanent. So the current situation is only slightly better than the one of Boolean circuit complexity, where nobody knows how to prove a lower bound on the complexity of a concrete Boolean function $\{0, 1\}^n \to \{0, 1\}$, which is superlinear in $n$.

At the end of the nineties, yet another algebraic theory of NP-completeness was proposed by Blum, Shub, and Smale [BSS89], with the intention of developing an appropriate complexity theory for numeric computation. This BSS-model received quite a lot of attention (see [BCSS98]), however, the separation of P and NP over the field of real or complex numbers also remained elusive.

The enormous progress in computational complexity was made possible by the introduction of complexity classes and the invention of ingenious reductions between problems. The significance of Valiant's contribution [Val79a] was to introduce these concepts in algebraic complexity, which has fuelled the progress of the field. His algebraic reformulation of the P versus NP as the permanent versus determinant problem has opened up a connection to established parts of mathematics that should be significant for the solution of the major questions in complexity theory.

**Acknowledgments.** I am grateful to Christian Ikenmeyer, Pascal Koiran, Mrinal Kumar and Rocco Servedio for providing detailed comments and pointing out several valuable references.

## 2. Foundations

We define and explain here the fundamental algebraic complexity classes and completeness results conceived by Valiant.

### 2.1. Arithmetic circuits.

The model of arithmetic circuits is the most common and general one to formalize the computations of polynomials over a fixed field $\mathbb{F}$ and some set $X$ of variables. It has the following short and clean definition.

**Definition 2.1.** An *arithmetic circuit* $\Phi$ over the field $\mathbb{F}$ and the set $X$ of variables is a directed acyclic graph with the following properties.

- Every node $v$ of $\Phi$ either has indegree zero or indegree two.
- Each node $v$ of indegree zero carries as a label either a variable from $X$ or an element of $\mathbb{F}$.
- Each node $v$ of indegree two carries the label $+$ or $\times$.

It is convenient to introduce some related terminology. Instead of nodes we also speak of gates. An input gate is one of indegree zero, the other gates are called operations gates: they are either addition or multiplication gates. The *size* $|\Phi|$ of $\Phi$ is defined as the number of its operation gates. The depth of a gate $v$ is defined as the maximal length of a directed path ending in $v$. The *depth* $D(\Phi)$ of $\Phi$ is the maximum depth of a gate of $\Phi$.

To a gate $v$ we assign the *subcircuit* $\Phi_v$ with root $v$, which is defined on the induced digraph consisting of the nodes $u$ for which there is a directed path to $v$. We denote by $X_v$ the set of variables occuring in $\Phi_v$, that is, $x \in X_v$ if $x \in X$ is the label of an input gate of $\Phi_v$. We define the *(formal) degree* $\deg(v)$ and the *polynomial* $\hat{\Phi}_v$ *computed by* $\Phi$ *at* $v$ by induction on the depth of $v$: if $v$ is an input gate with label $\alpha$, then $\hat{\Phi}_v := \alpha$ and $\deg(v) := 0$ if $\alpha \in \mathbb{F}$ and $\deg(v) := 1$ if $\alpha \in X$, if $v$ is an addition gate with children[2] $v_1, v_2$, then $\deg(v) := \max\{\deg(v_1), \deg(v_2)\}$ and $\hat{\Phi}_v := \hat{\Phi}_{v_1} + \hat{\Phi}_{v_2}$, and if $v$ is a multiplication gate with children $v_1, v_2$, then $\deg(v) := \deg(v_1) + \deg(v_2)$ and $\hat{\Phi}_v := \hat{\Phi}_{v_1} \cdot \hat{\Phi}_{v_2}$. Note that alway $\deg \hat{\Phi}_v \leq \deg(v)$. The *degree* $\deg \Phi$ of $\Phi$ is defined as the maximum of the $\deg(v)$. The circuit $\Phi$ is called *homogeneous* if all the $\hat{\Phi}_v$ are homogeneous polynomials. Finally, we say that $\Phi$ computes a polynomial $f$ if $f = \hat{\Phi}_v$ for some $v$. See Fig. 1 for an illustration.

Homogeneous circuits are much easier to control. By the following simple and fundamental result, we can assume that the circuit is homogenous, at least when focusing on moderate degrees (e.g., see [BCS97, §7.1]).

**Proposition 2.2.** *Let $\Phi$ be an arithmetic circuit $\Phi$ of size $s$ and let $d \geq 1$. Then there is a homogenous arithmetic circuit of size $O(sd^2)$, which computes all the homogeneous components of all the polynomials $\hat{\Phi}_v$ computed by $\Phi$.*

One calls $\Phi$ an *arithmetic formula* if the underlying digraph is a directed binary tree. This means that each gate has outdegree at most one. The idea behind this restriction is that results of intermediate computations can be used at most once. By induction one shows that $|\Phi| \leq 2^{D(\Phi)} - 1$ and equality holds if the underlying digraph is a complete binary tree. A remarkable result due to Brent [Bre75] states that every polynomial can be computed by an arithmetic formula of small depth.

---

[2] We say that a gate $v_1$ is a child of the gate $v$ if there is an edge from $v_1$ to $v$.

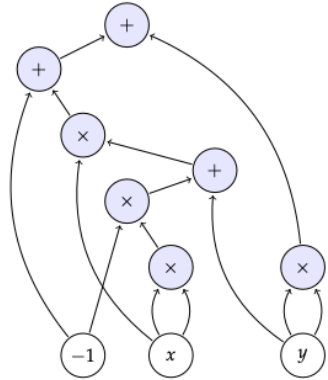

FIGURE 1. An arithmetic circuit of size 7, depth 6 and degree 3 computing $-x^3 + xy + y^2 - 1$.

**Theorem 2.3.** *Suppose the polynomial $f$ is computed by the arithmetic formula $\Phi$ of size $s$. Then there exists an arithmetic formula of depth at most $O(\log s)$ computing $f$.*

The following two complexity notions have been extensively studied.

**Definition 2.4.** The *complexity $L(f)$* of a polynomial $f$ is the minimal size of an arithmetic circuit computing $f$. The *expression size $E(f)$* is the minimal size of an arithmetic formula computing $f$.[3]

**Example 2.1.** 1. For the sum $f = x_1 + \cdots + x_n$ and product $f = x_1 \cdots x_n$ we have $L(f) = E(f) = n-1$. Indeed, the upper bounds are trivial and it is easy to see that $E(f) \geq L(f) \geq n-1$ for any polynomial $f$ depending on $n$ variables.

2. To illustrate the difference between expression size and complexity, we note that $E(x^d) = d - 1$ and $L(f) = \Theta(\log_2 d)$. Indeed, one easily proves that $E(f) \geq \deg f - 1$ and $L(f) \geq \log_2 \deg f$ by tracing the growth of the degrees of polynomials in a computation.

Since it turned out to be very hard to prove lower bounds in this general model, various restrictions of it have been studied (see Section 5).

As a first step towards this, we introduce the following syntactic restriction for controlling degrees from [Mal03, MP08].

**Definition 2.5.** An arithmetic circuit $\Phi$ is called *multiplicatively disjoint* if, for each multiplication gate $v$ with children $v_1, v_2$, the subcircuits $\Phi_{v_1}$ and $\Phi_{v_2}$ are disjoint.

By induction one shows that $\deg(v) \leq |\Phi_v| + 1$. Hence the size of a multiplicatively disjoint circuit provides a bound on its degree and hence on the degrees of the polynomials computed by it. We also have the following result from [MP08, Lemma 2].

**Proposition 2.6.** *Let $\Phi$ be an arithmetic circuit computing $f$. Then there is a multiplicatively disjoint arithmetic circuit $\Psi$ of size at most $O(|\Phi| \deg \Phi)$ computing $f$.*

---

[3]Often, the complexity is denoted $S(f)$ and the expression size $L(f)$; we stick here to Strassen's choice of notation.

Originally, Valiant [Val79a] used the model of arithmetic formulas of quasipolynomial size[4] to formulate a completeness result for the determinant. When focusing on circuits of polynomial size, it is more natural to relax the syntactic restriction to formulas. This leads to the notion of *weakly skew circuits*, which was first introduced by Toda [Tod92]. Unfortunately, this paper was not widely known, so that the notion was rediscovered much later by Malod [Mal03, MP08] in his thesis.

**Definition 2.7.**  (1) . An arithmetic circuit $\Phi$ is called *skew* if for each multiplication gate, one of its children is an input gate.

(2) An arithmetic circuit $\Phi$ is called *weakly-skew* if each multiplication gate $v$ has a distinguished son $v'$ with the property that $(v', v)$ is the only edge connecting a gate of the subcircuit $\Phi_{v'}$ with a gate not belonging to $\Phi_{v'}$.

*Remark* 2.8. 1. The idea behind weakly-skew circuits is that none of the intermediate results computed by the subcircuit $\Phi_{v'}$ may be reused by $\Phi$, if $v$ is multiplication gate. Clearly, a formula is weakly-skew. Note that weakly-skew circuits are multiplicatively disjoint, but the converse is not true. Moreover, by copying input gates, skew circuits may be assumed to be weakly-skew. We will see in Remark 2.22 that skew circuits have the same power as weakly ones.

2. The computation of the entries of the *iterated matrix product* $A_1 \cdots A_d$ from the entries of given $n \times n$ matrices $A_i$, based on the associativity relation $A_1 \cdots A_d = BA_d$ with $B = A_1 \cdots A_{d-1}$, can be done by a skew circuit of size $O(dn^3)$. The reason is that for computing the products $B_{ij}(A_d)_{j\ell}$, the circuit can directly access the input variables $(A_d)_{j\ell}$.

2.2. **Valiant's complexity classes.** The objects of interest are sequences $(f_n)$ of multivariate polynomials over a fixed field $\mathbb{F}$. A family of great interest is provided by the sequence of determinants,

$$\text{(2.1)} \qquad \text{DET}_n := \sum_{\pi \in S_n} \text{sgn}(\pi) \prod_{i=1}^{n} x_{i, \pi(i)}$$

of a square matrix of size $n$ having distinct variables $X_{ij}$ as entries. This sequence encapsulates the complexity of linear algebra, as far as algebraic computation is concerned. In this example, the $f_n$ are strongly related. However, for the general definition, it turned out to be convenient not to assume any relationship between the $f_n$: the model is "nonuniform".[5]

By putting various restrictions on the structure of polynomial size arithmetic circuits $\Phi_n$ computing $f_n$, we arrive at the definitions of Valiant's complexity classes.

**Definition 2.9.**  (1) The complexity class $\text{VP}^{\mathbb{F}}$ is defined as the set of sequences $(f_n)_{n \in \mathbb{N}}$ of multivariate polynomials over $\mathbb{F}$ such that there exists a sequence $(\Phi_n)_{n \in \mathbb{N}}$ of multiplicatively disjoint arithmetic circuits over $\mathbb{F}$ such that $\Phi_n$ computes $f_n$ and the size $|\Phi_n|$ is bounded by a polynomial in $n$.

(2) If we require the circuits $\Phi_n$ to be weakly skew, we obtain the class $\text{VBP}^{\mathbb{F}}$.

(3) If we make the even stronger assumption that the $\Phi_n$ are formulas, we obtain the class $\text{VF}^{\mathbb{F}}$.

Usually one drops the dependence on the given field $\mathbb{F}$ for notational simplicity. Let $X_n$ be the set of variables of $f_n$. Often, one thinks of $X_n \subseteq X_{n+1}$, but since we do not require any relationship among the $f_n$, no relationship between the $X_n$ is required. We have the inclusions of complexity classes

$$\text{(2.2)} \qquad \text{VF} \subseteq \text{VBP} \subseteq \text{VP}$$

which are believed to be strict. The class VBP, which may seem artifical at first, derives its relevance from the insight that $(\text{DET}_n)$ can be identified as a complete problem in VBP, see Section 2.5.

---

[4]Size bounded by $2^{\log^c n}$.

[5]This assumption is not of great relevance and was made by Valiant for the sake of elegance and convenience.

*Remark* 2.10. 1. By Proposition 2.6, the class VP can also be characterized as the set of sequences $(f_n)$ of polynomials such that $\deg f_n$ and $L(f_n)$ grow at most polynomially in $n$.

2. The requirement of the multiplicative disjointness is a strong one: the Boolean analogue of the class VP is the class LOGCFL, see [Ven91]. (Without it, VP would correspond to the class P of polynomial time.)

3. Different names for these classes are in use. The name VBP derives from algebraic *branching programs* (see Section 2.5), while VF relates to *formulas*. In the literature, VBP is sometimes denoted $VP_{ws}$ for "weakly skew" while VF is denoted $VP_e$ for "expression". We believe the above choice to be simpler and more uniform. By Remark 2.22 below, VBP can also be defined in terms of skew circuits.

4. The thesis by Mengel [Men13] gives a nice explanation of the difference in power between the classes VP and VBP in terms of stack algebraic branching programs.

The parallel complexity classes $VNC_k^{\mathbb{F}}$ are subclasses of VP obtained by putting depth restrictions on the circuits as follows ($k$ is a positive integer).[6]

**Definition 2.11.** The complexity class $VNC_k^{\mathbb{F}}$ is defined as the set of sequences $(f_n)_{n \in \mathbb{N}}$ of multivariate polynomials over $\mathbb{F}$ such that there exists a sequence $(\Phi_n)_{n \in \mathbb{N}}$ of multiplicatively disjoint arithmetic circuits over $\mathbb{F}$ such that $\Phi_n$ computes $f_n$, the size $|\Phi_n|$ is bounded by a polynomial in $n$, and the depth satisfies $D(\Phi_n) = O(\log^k n)$.

Surprisingly, the hierarchy $VNC_1 \subseteq VNC_2 \subseteq \ldots \subseteq VP$ collapses. This is due to the following important general parallelization result due to Valiant and Skyum [SV83], which was simplified and improved in Valiant et al. [VSBR83], who proved the following.

**Theorem 2.12.** *Suppose a polynomial $f$ of degree $d$ is computed by an arithmetic circuit of size $s$. Then $f$ can be computed by an arithmetic circuit of size $O(d^6 s^3)$ and depth $O(\log(d) \log(sd))$.*

The first step of the proof is to assume that the circuit is homogeneous, which is possible by Proposition 2.2. By focusing on the degree one can construct a balanced circuit by a sophisticated procedure. We refer to [RY08] for a comprehensive proof of Theorem 2.12 in somewhat more generality.

**Corollary 2.13.** *1. Suppose a polynomial $f$ is computed by an arithmetic circuit of depth $D$. Then $f$ can be computed by an arithmetic formula of the same depth.*

*2. Suppose a polynomial $f$ of degree $d$ is computed by an arithmetic circuit of size $s$. Then $f$ can be computed by an arithmetic formula of size $(sd)^{O(\log d)}$.*

*Proof.* 1. This is obtained by just recomputing results, ignoring a potential exponential increase in size.

2. Combine part one with Theorem 2.12. □

From this and Theorem 2.3 we easily obtain the following.

**Corollary 2.14.** *Over any field $\mathbb{F}$ we have*

(1) $VF = VNC_1$,
(2) $VNC_2 = VNC_3 = \cdots = VP$.

*Proof.* 1. To show the inclusion $VNC_1 \subseteq VF$, we transform circuits into formulas by preserving the depth (see Corollary 2.13(1)), and then use that $|\Phi| \leq 2^{D(\Phi)} - 1$ for every formula $\Phi$. The reverse inclusion is a consequence of Theorem 2.3.

2. This is a consequence of Theorem 2.12. □

---

[6]The terminology is borrowed from "Nick's class" as for the boolean classes.

2.3. **Reduction via substitution.** In order to identify complete problems in the above defined algebraic complexity classes, a notion of reduction is required. The choice made by Valiant [Val79a] stands out by its simplicity and elegance.

**Definition 2.15.** 1. Let $f(x_1, \ldots, x_n)$ and $g(y_1, \ldots, y_m)$ be multivariate polynomials over the field $\mathbb{F}$. We call $f$ a *projection* of $g$, written $f \leq g$, if there exists $a_i \in \mathbb{F} \cup \{x_1, \ldots, x_n\}$ such that $f(x_1, \ldots, x_n) = g(a_1, \ldots, a_m)$.

2. Let $(f_n)$ and $(g_m)$ be sequences of multivariate polynomials over $\mathbb{F}$. We call $(f_n)$ a *p-projection* of $(g_m)$ if there exists a polynomially bounded function $t \colon \mathbb{N} \to \mathbb{N}$ such that $f_n \leq g_{t(n)}$ for all $n$.

In other words, $f_n$ is obtained from $g_m$ by substituting the variables of $g_m$ by variables of $f$ and constants, where $m$ is upper bounded by a polynomial in $n$. This notion of reduction does not require any computation, so that it surprising that it turns out to be strong enough to arrive at meaningful completeness results.[7] This reduction notion also turns out to be useful in the Boolean setting [SV85].

It is straighforward to check from the definition that the classes introduced in Section 2.2 are closed under taking p-projections. Completeness is defined in the usual way.

**Definition 2.16.** Let $\mathscr{C}$ be a class of sequences of polynomials over $\mathbb{F}$. A family $(g_m)$ is called $\mathscr{C}$-complete if $(g_m) \in \mathscr{C}$ and if any $(f_n) \in \mathscr{C}$ is a p-projection of $(g_m)$.

It is easy to provide an example of a VF-complete sequence. Consider $g_n := \mathrm{trace}(M_1 \cdots M_n)$, where the $M_\nu$ are $3 \times 3$-matrices whose entries are different variables. Using $M_1 \cdots M_n = (M_1 \cdots M_{n-1})M_n$, we see that $g_n$ has expression size $O(n)$, thus $(g_n) \in$ VF. In order to prove the VF-completeness of $(g_n)$ it suffices to verify that any polynomial $f$ of expression size $m$ is a projection of $g_{t(m)}$, where $t(m) = O(m^2)$; see [BOC92].

2.4. **Circuits for the determinant.** It is a fundamental fact that determinants can be efficiently computed. In fact, this is the deeper reason that linear algebra is computationally feasible. Furthermore, it is fair to say that the bulk of everyday computation is spent on linear algebra calculations. However, looking at the definition (2.1) of the determinant $\mathrm{DET}_n$, it is not all obvious why it can be computed by arithmetic circuits of size polynomially bounded in $n$. The shortest way to see this is to apply Gaussian elimination, which takes $O(n^3)$ operations, but this requires divisions. By a general principle due to Strassen [Str73b], the divisions can be eliminated, resulting in a division-free arithmetic circuit of size $O(n^5)$. This shows that the sequence $(\mathrm{DET}_n)$ belongs to the class VP.

There are more direct ways to produce efficient division-free arithmetic circuits for the determinant, which also allow to design small depth circuits for it. It may not be surprising that some ideas for this task are quite old. Apparently, the oldest method goes back to 1840 and was invented by the astronomer Le Verrier [LV40], who became famous as one of the discoverer of the planet Neptune. Le Verrier's method computes the coefficients $c_1, \ldots, c_n$ of the characteristic polynomial

$$\det(tI_n - A) = t^n - c_1 t^{n-1} - c_2 t^{n-2} - \ldots - c_n$$

of a matrix $A \in \mathbb{F}^{n \times n}$. (Note $c_n = (-1)^n \det(A)$.) This is done by first computing the traces of matrix powers

$$s_1 := \mathrm{trace}(A), s_2 := \mathrm{trace}(A^2), \ldots, s_n := \mathrm{trace}(A^n),$$

---

[7]From the viewpoint of projective algebraic geometry, it is more natural to substitute variables by scalar multiples of variables, see [DGI$^+$24b].

and then solving the linear system $S(c_1, \ldots, c_n)^T = (s_1, \ldots, s_n)^T$ for $c_1, \ldots, c_n$ by inverting the triangular matrix (Newton relations)

$$S = \begin{bmatrix} 1 & 0 & \ldots & \ldots & 0 \\ s_1 & 2 & \ddots & & \vdots \\ \vdots & \ddots & \ddots & \ddots & \vdots \\ s_{n-2} & s_{n-3} & & \ddots & 0 \\ s_{n-1} & s_{n-2} & \ldots & s_1 & n \end{bmatrix}.$$

This method translates to an arithmetic circuit of size $O(n^4)$ computing $c_1, \ldots, c_n$ from the entries of $A$. However, this works only over fields of characteristic zero since the division by the determinant $n! = \det(S)$ is required. It is not hard to parallelize this approach in order to arrive at an arithmetic circuit of size $O(n^4)$ and depth $O(\log^2 n)$ (Csanky [Csa76]). For this, one computes the matrix powers in parallel and solves the triangular linear system recursively, based on

$$S = \begin{bmatrix} S_1 & 0 \\ S_3 & S_2 \end{bmatrix}, \quad S^{-1} = \begin{bmatrix} S_1^{-1} & 0 \\ -S_2^{-1} S_3 S_1^{-1} & S_2^{-1} \end{bmatrix}.$$

(One may assume that $n$ is a power of 2 by enlarging $n$.) A more detailed inspection reveals that the resulting arithmetic circuit can be assumed to be skew. Indeed, most of the work reduces to computing iterated matrix products.

There are division-free efficient algebraic methods for computing the coefficients of the characteristic polynomial that work over an arbitrary commutative ring and translate to (weakly skew) arithmetic circuits of size $O(n^4)$ and depth $O(\log^2 n)$. Berkowitz [Ber84] found such a method relying on ideas of Samuelson [Sam42] and a different method is due to Chistov [Chi85]. (We refer to [AL04] for a comprehensive account.) Yet another, completely combinatorial approach for computing the determinant is due to Mahajan and Vinay [MV97].

We summarize these finding as follows.

**Proposition 2.17.** *There is a sequence of weakly-skew arithmetic circuits of size $O(n^4)$ and depth $O(\log^2 n)$ which computes $\mathrm{DET}_n$ over any commutative ring, using integer coefficients. In particular, we have $(\mathrm{DET}_n) \in \mathrm{VBP}$ over any field.*

By recomputing intermediate results, a circuit $C$ of depth $D$ can be transformed to a formula $\Phi$ of the same depth $D$, see Corollary 2.13. The size of $\Phi$ is upper bounded by $2^D - 1$. Therefore, Proposition 2.17 implies:

**Corollary 2.18.** *There is a sequence of arithmetic formulas of size $2^{O(\log^2 n)}$ which computes $\mathrm{DET}_n$ (over any commutative ring, using integer coefficients).*

It is an intriguing open question whether $\mathrm{DET}_n$ can be computed by an arithmetic formula of size polynomially bounded in $n$. In fact:

**Corollary 2.19.** *The following statements are equivalent:*

(1) $\mathrm{DET}_n$ *can be computed by a sequence of arithmetic formulas of polynomial size,*
(2) $\mathrm{DET} \in \mathrm{VF}$,
(3) $\mathrm{VF} = \mathrm{VBP}$.

*Proof.* (1)$\Rightarrow$(2): follows from Theorem 2.3.
(2)$\Rightarrow$(3): follows from Theorem 2.20 below.
(3)$\Rightarrow$(1): clear. $\qquad \square$

Kalorkoti [Kal85a] proved that the computation of $\mathrm{DET}_n$ requires arithmetic formulas of cubic size: $E(\mathrm{DET}_n) = \Omega(n^3)$. His proof technique is an adaptation of a lower bound method for Boolean functions due to Nečiporuk [Neč66].

### 2.5. Completeness of the determinant.

Our goal is to prove the following sharper version [Tod92, MP08] of Valiant's completeness result for the determinant from [Val79a].

**Theorem 2.20.** *The sequence* $(\mathrm{DET}_n)$ *is* VBP-*complete over any field.*

Let us point out that, combined with Proposition 2.17, this implies a weaker version of Theorem 2.12: for $(f_n) \in \mathrm{VBP}$, there is a sequence of weakly-skew arithmetic circuits of size $n^{O(1)}$ and depth $O(\log^2 n)$ computing $f_n$.

The beautiful proof of Theorem 2.20 proceeds in two steps, which we now outline. We first introduce the model of *algebraic branching programs* ABPs, which consists of an acylic digraph $G = (V, E)$ with two distinguished nodes $s, t$ (source and sink), and a weight function $w\colon E \to \mathbb{F} \cup X$, where $X$ is a set of variables. Parallel edges are allowed. (Often, one requires the graph to be layered, having only edges between consecutive layers; see [BIM$^+$20].) By an $s$-$t$-path $\pi$ we understand a simple directed path between $s$ and $t$. The weight $w(\pi)$ of an $s$-$t$-path $\pi$ is defined as the product of the weights of the edges of $\pi$. By definition, the ABP *computes* the value $\mathrm{SW}(G)$ defined as the sum of the weights of all $s$-$t$-paths of $G$.

**Proposition 2.21.** *Let* $\Phi$ *be a weakly-skew arithmetic circuit of size $m$ computing $f$. Then there exists an ABP computing $f$ with at most $m + 2$ nodes and at most $m + 1$ edges.*

*Proof.* We only describe the proof in the special case where $\Phi$ is a formula, that is, the underlying graph is a directed tree. The proof of the general case proceeds similarly, but requires some more care.

We proceed by induction on $m$. If $m = 0$, then $f \in \mathbb{F} \cup X$ and the assertion is trivially satisfied.

Assume that $f$ is computed at the gate $v$: without loss of generality it is not an input gate. Applying the induction hypothesis to the subcircuits $\Phi_{v_i}$, for $i = 1, 2$, we see that there exist formulas $G_1$ and $G_2$ such that $\hat{\Phi}_{v_i} = \mathrm{SW}(G_i)$. Note that $|\Phi_v| = |\Phi_{v_1}| + |\Phi_{v_2}| + 1$.

Assume first that $v$ is an addition gate with children $v_1$ and $v_2$. We take the disjoint union of $G_1$ and $G_2$ and identify the sources and sinks of the $G_i$, respectively. Then the resulting digraph $G = (V, E)$ clearly satisfies $\mathrm{SW}(G) = \mathrm{SW}(G_1) + \mathrm{SW}(G_2) = \hat{\Phi}_{v_1} + \hat{\Phi}_{v_2} = \hat{\Phi}_v = f$. Moreover, by induction hypothesis, $|V| = |V_1| + |V_2| - 2 \le |\Phi_{v_1}| + 2 + |\Phi_{v_2}| + 2 - 2 = |\Phi_v| + 1$ and $|E| = |E_1| + |E_2| \le |\Phi_{v_1}| + 1 + |\Phi_{v_2}| + 1 = |\Phi_v| + 1$ and $|\Phi_v| \le m$.

Assume now that $v$ is a multiplication gate. We take disjoint union of the digraphs $G_1$ and $G_2$ and identify the sink of $G_1$ with the source of $G_2$. An $s$-$t$-path $\pi$ of the resulting digraph $G$ is given by a pair $\pi_1, \pi_2$ of $s$-$t$-paths of $G_1, G_2$, respectively, and therefore $w(\pi) = w(\pi_1)w(\pi_2)$. This implies $\mathrm{SW}(G) = \mathrm{SW}(G_1)\mathrm{SW}(G_2) = \hat{\Phi}_{v_1}\hat{\Phi}_{v_2} = \hat{\Phi}_v = f$. Moreover, $|V| = |V_1| + |V_2| - 1 \le |\Phi_v| + 2$ and again $|E| = |E_1| + |E_2| \le |\Phi_v| + 1$. $\qquad\square$

*Remark* 2.22. It is easy to check that a computation of $f$ by an ABP with $m + 2$ nodes can be turned into a skew circuit of size $m$ computing $f$. Together with Proposition 2.21, we see that if $f$ can be computed by a weakly skew circuit of size $m$, then it can also be computed by a skew circuit of size $m$.

Theorem 2.20 follows from the following result.

**Proposition 2.23.** *Suppose $f$ can be computed by a weakly-skew arithmetic circuit of size $m$. Then $f$ is a projection of* $\mathrm{DET}_{m+1}$. *More specifically, $f = \det(A)$ for some sparse matrix $A$ of side length $m+1$ with entries in $\mathbb{F} \cup X$. Sparse means that outside its main diagonal, $A$ has at most $m + 1$ nonzero entries. Moreover, the entries on the main diagonal are either $0$ or $1$.*

*Proof.* A key idea is to view matrices as weighted adjacency matrices of digraphs.

Suppose $H$ is an edge weighted digraph. After ordering the set of nodes of $H$, we can identify $H$ with the matrix $A$, where $A_{ij}$ is the weight of the edge pointing from the $i$th node to the $j$th node. (If there is no such edge, we set $A_{ij} = 0$.) A permutation of the nodes of $H$ has a unique decomposition into cycles and can thus be identified and visualized as a *cycle cover* of the digraph $H$: this is a set of directed cycles of $H$ whose node sets form a partition of the node set of $H$. We define the weight $w(\sigma)$ of a cycle cover $\sigma$ as the product of the weights of the edges occuring in $\sigma$. Using this notation, we can write the determinant of $A$ as the sum of the weights of all cycle covers $\sigma$ of $H$, multiplied with the corresponding signs:

$$\det(A) = \sum_\sigma \operatorname{sgn}(\sigma) w(\sigma).$$

Let $G = (V, E)$ denote the digraph underlying the algebraic branching program $\Phi$ computing $f$ resulting from Proposition 2.21. We have $|V| \le m + 2$ and $|E| \le m + 1$. We replace each weight $w(e)$ by $\tilde{w}(e) := -w(e)$ for $e \in E$. Then we identify the source with the target and we introduce loops of weight 1 at all nodes different from this node. Let us denote the resulting edge weighted digraph by $H$. Then $|H| = |G| - 1 \le m + 1$ and $H$ has at most $|E| \le m + 1$ many edges, which are not loops.

To any $s$-$t$-path $\pi$ of $G$ there corresponds a cycle cover $\sigma(\pi)$ of $H$ consisting of loops and of the cycle obtained from $\pi$ after identifying $s$ with $t$. This correspondence is bijective since $G$ is acyclic. Moreover,

$$\tilde{w}(\sigma(\pi)) = (-1)^{\ell(\pi)} w(\pi),$$

where $\ell(\pi)$ equals the number of edges of the path $\pi$. Using $\operatorname{sgn}(\pi) = (-1)^{\ell(\pi)-1}$, we obtain for the sum over all cycle covers $\sigma$ of $H$:

$$\sum_\sigma \operatorname{sgn}(\sigma) \tilde{w}(\sigma) = \sum_\pi \operatorname{sgn}(\sigma(\pi))(-1)^{\ell(\pi)}(\pi) w(\pi) = -\sum_\pi w(\pi) = -f.$$

Thus, if $A$ denotes the adjacency matrix of $H$ (fixing some ordering of the nodes of $H$), we have $\det(A) = -f$. We can remove the sign by exchanging two columns of $A$. Finally, note that $A$ is a matrix of format $m + 1$ and satisfies the sparsity condition. $\qquad\square$

We remark that for his completeness results, Valiant [Val79a] used the more generous notion of $qp$-reduction, only requiring that $t(n)$ is quasipolynomially bounded in $n$, that is, $t(n) \le 2^{\log^c n}$ for some constant $c$.

Consider the iterated matrix multiplication family defined by $\mathrm{IMM}_{n,d} := \operatorname{trace}(A_1 \cdots A_d)$, where the $A_i$ are $n \times n$ matrices with independent indeterminate entries. We already noted in Remark 2.8 that $\mathrm{IMM} \in \mathrm{VBP}$. Sometimes, this family is instead defined as the $(1,1)$-entry of the matrix product, which destroys its invariance. However, this does not affect the completeness result below, which is obtained by a modification of the proof of Proposition 2.23.

**Corollary 2.24.** *The iterated matrix multiplication family* $(\mathrm{IMM}_{n,n})$ *is* VBP-*complete over any field.*

Inspired by [Coo85], more natural VBP-families can be identified. However, the question to identify natural complete problems for the class VP was open for a long time. Finally, a first solution in terms of the polynomial enumerators of graph homomorphisms was found in [DMM+16] and improved in [MS18].

2.6. **The class** VNP. In his seminal work [Val79a], Valiant came up with the following algebraic analogue of the complexity class NP.

**Definition 2.25.** A sequence $(f_n)$ of multivariate polynomials over the field $\mathbb{F}$ is called *p-definable* iff there exists a sequence $g = (g_n)$ in $\mathrm{VP}^{\mathbb{F}}$, $g_n \in \mathbb{F}[x_1, \ldots, x_{u(n)}]$, such that for all $n$

$$f_n(x_1, \ldots, x_{v(n)}) = \sum_{e \in \{0,1\}^{u(n)-v(n)}} g_n(x_1, \ldots, x_{v(n)}, e_{v(n)+1}, \ldots, e_{u(n)}) \ .$$

The set of *p-definable* sequences form the complexity class $\mathrm{VNP}^{\mathbb{F}}$ over $\mathbb{F}$.

The sequences $(g_n)$ in VP were called *p-computable* in [Val82] and thought to model the feasible problems. By contrast, the members $f_n$ of a *p-definable* sequence are obtained by a summation of exponentially many values of $g_n$. Therefore, the complexity of $f_n$ is expected to be exponential in $n$.

We have the obvious inclusion $\mathrm{VP}^{\mathbb{F}} \subseteq \mathrm{VNP}^{\mathbb{F}}$ and Valiant conjectured these classes to be different.

**Valiant's conjecture.** We have $\mathrm{VP}^{\mathbb{F}} \neq \mathrm{VNP}^{\mathbb{F}}$ over any field $\mathbb{F}$.

This fundamental conjecture is widely open and considered the holy grail of algebraic complexity theory. Strassen [Str87b] had named it "Valiant's Hypothesis" and pointed out the analogy with the conjecture $\mathrm{P} \neq \mathrm{NP}$ due to Cook [Coo71].

*Remark* 2.26. 1. It is obvious that the definition of VNP is closely related to Valiant's counting complexity class #P capturing the complexity of counting problems [Val79b, Val79c], rather than to the class NP. (In that sense the choice of name is a bit unfortunate.)

2. In [Bür00b] it was shown that the truth of $\mathrm{VP}^{\mathbb{F}} \neq \mathrm{VNP}^{\mathbb{F}}$ only depends on the characteristic of the field $\mathbb{F}$.

3. The known logical relations of Valiant's conjecture to other conjectures are discussed in the Sections 4.4–4.6.

4. Mengel [Men13] characterizes VNP by a notion of "random access algebraic branching programs".

An important example of a *p-definable* family is provided by permanents. The *permanent* $\mathrm{per}(A)$ of an $n$ by $n$ matrix $A = (a_{i,j})$ is defined as

$$\mathrm{per}(A) = \sum_{\pi \in S_n} \prod_{i=1}^n a_{i,\pi(i)}.$$

This matrix function has a nice combinatorial interpretation: the permanent of the adjacency matrix of a bipartite graph equals the number of perfect matchings of this graph. The permanent is the object of intense studies in combinatorics [Min78].

The fastest known algebraic algorithms for computing the permanent rely on the following "inclusion-exclusion" formula due to Ryser [Rys63]:

$$(2.3) \qquad \mathrm{per}(A) = \sum_I (-1)^{|I|} \prod_{i=1}^n \Big( \sum_{j \notin I} a_{i,j} \Big),$$

where the sum is over all subsets $I$ of $\{1, \ldots, n\}$.

We denote by $\mathrm{PER}_n$ the permanent of an $n$ by $n$ matrix of independent variables over some fixed field $\mathbb{F}$. Ryser's formula shows $E(\mathrm{PER}_n) = O(n^2 2^n)$. It is unknown whether $(\mathrm{PER}_n) \in \mathrm{VP}$. On the other hand, $(\mathrm{PER}_n) \in \mathrm{VNP}$, but checking this directly is a bit cumbersome. However, this immediately follows from the following general principle due to Valiant [Val79a], which links the class VNP to its Boolean counterpart #P (see [Bür00a, Prop. 2.20]).

**Proposition 2.27** (Valiant's Criterion). *Suppose* $\phi \colon \{0,1\}^* \to \mathbb{N}$ *is a function in the nonuniform class* #P/poly.[8] *Then the family* $(f_n)$ *of polynomials defined by*

$$f_n = \sum_{e \in \{0,1\}^n} \phi(e) x_1^{e_1} \cdots x_n^{e_n}$$

---

[8] The class #P with polynomial advice.

*is p-definable.*

The easy proof relies on simulating Boolean operations by arithmetic operations. (Note that the definition #P involves counting satisfying assignements of Boolean formulas in conjunctive normal form.) The proof of the following result from [Val82] relies on similar ideas.

**Proposition 2.28.** *Definition 2.25 of the class* VNP *does not change when making the stronger requirement* $(g_n) \in$ VF.

*Proof.* (Sketch) It is enough to construct a representation with $(g_n) \in$ VF for sequences $(f_n) \in$ VP. Fix $n$ and suppose $C$ is a multiplicatively disjoint arithmetic circuit computing $f := f_n$. We denote by $G = (V, E)$ the underlying digraph. Further, we denote by $I \subseteq V$ the set of input gates of $C$ and by $\lambda_i \in \mathbb{F} \cup X$ the label at $i \in I$.

The key idea is to write $f = \sum_T \mathrm{val}(T)$, where the sum is over all "parse trees" of $C$. The latter are directed trees formed from $C$, which record the generation of the monomials of $f$ (see [MP08] for the definition). Hereby, certain parts of $C$ may be copied several times, taking into account the fact that a circuit may reuse intermediate results. The monomial $\mathrm{val}(T)$ is defined as the product of the labels of the input gates of $T$.

One can encode the subgraphs of $G$ by indicators $(p, a) \in \{0, 1\}^V \times \{0, 1\}^E$, where $p(v) = 1$ expresses that node $v$ is selected and $a(u, v) = 1$ means that edge $(u, v)$ is selected. The fact that the subgraph is a parse tree of $C$ can then be expressed by a Boolean formula of roughly the size of $C$. We can simulate the Boolean formula by an arithmetic formula $\varphi(p, a)$ of roughly the same size. This way, one arrives at a representation

$$f = \sum_{p,a} \varphi(p, a) \prod_{i \in I} (p_i \lambda_i + 1 - p_i),$$

where the sum is over all binary $(p, a) \in \{0, 1\}^V \times \{0, 1\}^E$. The summand has expression size polynomially bounded in the size of $C$. $\square$

2.7. **Completeness of the permanent.** The following central theorem due to Valiant [Val79a, Val82] shows that PER is VNP-complete. (Note that permanents and determinants coincide in characteristic two.)

**Theorem 2.29.** *Over fields of characteristic different from two, the sequence* PER *is* VNP-*complete.*

This result immediately follows together with Proposition 2.28 from the following key ingredient.

**Proposition 2.30.** *Assume the polynomial* $g \in F[x_1, \ldots, x_n, y_1, \ldots, y_m]$ *can be computed by a formula of size less than $s$ and put*

$$f(x) = \sum_{e \in \{0,1\}^m} g(x, e).$$

*Then $f$ is a projection of* $\mathrm{PER}_{6s}$.

The proof of this result is based on an ingenious gadget construction. As for Theorem 2.20, it is important to think of a matrix $A$ as the weighted adjacency matrix of a digraph $G$. We thus define the permanent $\mathrm{per}(G)$ of $G$ as the permanent of the matrix $A$. (This is well defined, since the permanent is invariant under permutations of rows or columns.)

Suppose we have an edge weighted digraph $G$ with distinguished edges $c = (u, v)$ and $c' = (u', v')$. We assume that $u, v, u', v'$ are pairwise distinct. We insert between $c$ and $c'$ an auxiliary digraph $K$ according to Figure 2. (The edges $c_- := (u, p_1)$ and $c'_- := (u', p_3)$ are given the weights $w(c)$ and $w(c')$, respectively; edges have weight 1 unless otherwise indicated.) That is, we introduce two additional nodes $p_1, p_2$ on $c$ and one additional node $p_3$ on $c'$ and connect the nodes $p_i$ according to the following weight matrix (also denoted by $K$)

$$K := \begin{pmatrix} -1 & 1 & \frac{1}{2} \\ 1 & 1 & -\frac{1}{2} \\ 1 & 1 & -\frac{1}{2} \end{pmatrix} .$$

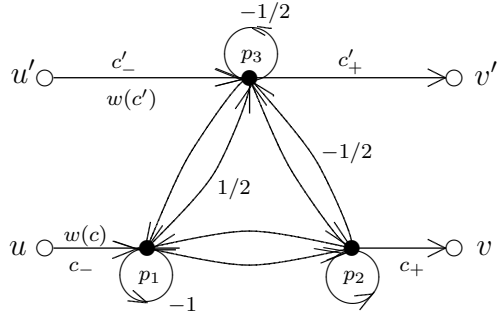

FIGURE 2. Iff-coupling of the edges $c = (u, v)$ and $c' = (u', v')$.

The resulting digraph is denoted by $\overline{G}$. We will say that $\overline{G}$ results from $G$ by *iff-coupling* of the edges $c$ and $c'$. The next lemma explains this naming.

**Lemma 2.31.** *The permanent of $\overline{G}$ equals the sum of the weights of all cycle covers of $G$, which contain either both of the edges $c$ and $c'$, or none.*

The proof is a straightforward verification based on

$$(2.4) \quad \operatorname{per}(K[2|1]) = \operatorname{per}(K[2|3]) = \operatorname{per}(K[3|1]) = \operatorname{per}(K[3|3]) = 0, \ \operatorname{per}(K) = \operatorname{per}(K[2,3|1,3]) = 1,$$

where $K[R|C]$ denotes $K$ with rows in $R$ and columns in $C$ removed.

*Remark* 2.32. Looking for a matrix $K$ satisfying (2.4) leads to a system of polynomial equations. If the corresponding system with per replaced by det, had a complex solution, then an adaptation of the proof would show that VP = VNP. Of course, this system does not have a solution. The observation to design clever algebraic gadgets led Valiant to the construction of holographic algorithms [Val08]. This has resulted in a fertile stream of works by Cai and his coauthors, see [CL11] for more information and references.

The proof of Proposition 2.30 relies on further auxiliary digraphs: the *rosettes* $R(\mu)$ defined for a positive integer $\mu$ (cf. Figure 3). They are constructed as follows. We start with a directed cycle of length $\mu$, consisting of the nodes $u_1, \ldots, u_\mu$ and the *connector edges* $c_i = (u_i, u_{i+1})$. (Here and in the following, the index arithmetic is supposed to be modulo $\mu$.) We introduce additional nodes $v_1, \ldots, v_\mu$ together with the edges $(u_i, v_i)$, $(v_i, u_{i+1})$. Finally, we add loops $(u_i, u_i)$ and $(v_i, v_i)$ at all the nodes. The resulting digraph will be called the rosette $R(\mu)$. All edges of $R(\mu)$ are supposed to carry the weight 1. We denote by $C = \{c_1, \ldots, c_\mu\}$ the set of connector edges of $R(\mu)$. The rosette $R(\mu)$ has the following easily verified properties.

- For each nonempty subset $S \subseteq C$ there is exactly one cycle cover of $R(\mu)$, which, among the connector edges, contains exactly the edges in $S$.
- There are exactly two cycle covers of $R(\mu)$, which contain no connector edge. One of them consists of loops only.

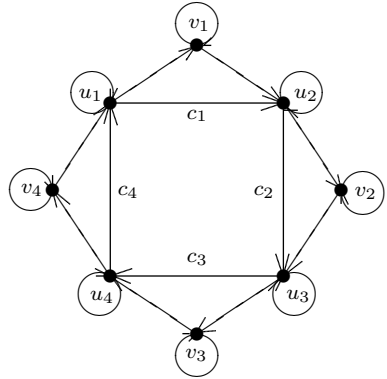

FIGURE 3. The rosette $R(4)$.

*Proof outline of Proposition 2.30.* Suppose $f(x_1, \ldots, x_n, y_1, \ldots, y_m)$ can be computed by a formula of size less than $s$. Similarly as for the proof of Proposition 2.23, we see that there is a digraph $G$ with $|G| \leq s$ such that $g = \operatorname{per}(G)$. The weights of the edges of $G$ are variables $x_i, y_j$ or constants in $\mathbb{F}$. Let $d_{i,1}, \ldots, d_{i,\mu_i}$ be the edges of $G$ carrying the weight $Y_i$. By the sparsity condition, we have $\mu_1 + \ldots + \mu_m \leq s$. We may assume that $\mu_i > 0$ for all $i$.

We form the disjoint union $F$ of the digraph $G$ and the rosettes $R(\mu_i)$ for $1 \leq i \leq m$. The connector edges of $R(\mu_i)$ are denoted by $c_{i,j}$ for $1 \leq j \leq \mu_i$. The digraph $F$ inherits its edge weights $w_F$ from $G$ and the rosettes, except that we give all the edges $d_{i,j}$ the new weight 1.

Let $\Phi$ denote the set of cycle covers of $F$ which, for all $i, j$, contain the edge $c_{i,j}$ iff they contain the edge $d_{i,j}$. Such cycle covers are said to respect iff-couplings of the edges $c_{i,j}$ with $d_{i,j}$. Based on the properties of the rosettes listed before, one verifies that

$$(2.5) \qquad f(x) = \sum_{\varphi \in \Phi} w_F(\varphi).$$

We can realize all the iff-couplings between $c_{i,j}$ and $d_{i,j}$ by means of the auxiliary construction of Figure 2. Let $\overline{F}$ be the edge weighted digraph thus resulting from $F$. By Lemma 2.31 we have $\operatorname{per}(\overline{F}) = \sum_{\varphi \in \Phi} w_F(\varphi)$. From (2.5) we can then conclude that $\operatorname{per}(\overline{F}) = f$. Moreover, we have

$$|F'| \leq |G| + 2\sum_{i=1}^{m} \mu_i + 3\sum_{i=1}^{m} \mu_i \leq 6s$$

and thus the proposition follows. □

In view of the completeness results for VBP (Theorem 2.20) and VNP (Theorem 2.29) it seems cleaner to focus on the conjecture VBP $\neq$ VNP, which is weaker than the conjecture VP $\neq$ VNP.

The separation of complexity classes VBP $\neq$ VNP can be rephrased as a quantitative, but otherwise purely algebraic comparison of the permanent and determinant.

**Corollary 2.33.** *Suppose* $\operatorname{char} \mathbb{F} \neq 2$. *Then showing* VBP $\neq$ VNP *is equivalent to proving that* PER *is not a p-projection of* DET.

This result is the starting point of geometric complexity theory, see Section 7.

2.8. **Generating functions of graph properties.** Related to $(\operatorname{PER}_n)$ is the sequence $(\operatorname{HC}_n)$ of *Hamilton cycle polynomials*, defined by

$$\operatorname{HC}_n = \sum_{\pi} \prod_{i=1}^{n} x_{i,\pi(i)},$$

where the sum is over all cycles $\pi \in S_n$ of length $n$. Note that the value of $\mathrm{HC}_n$ at the adjacency matrix of a digraph equals the number of its Hamilton cycles. Valiant [Val79a] proved that this sequence is VNP-complete *over any field* $\mathbb{F}$. Plenty of sequences $(f_n)$ of polynomials have been identified as being VNP-complete, see [Bür00a, Chap. 3]. Of course, Valiant's conjecture is equivalent to proving $(f_n) \notin \mathrm{VP}$ for any such VNP-complete sequence.

There are general ways to associate polynomials with a property $\mathscr{E}$ of graphs. Consider a graph $G = (V, E)$ with an edge weight function $w \colon E \to I := \mathbb{F} \cup X$, where again $X$ denotes a set of variables over a field $\mathbb{F}$. We extend the weight function $w$ to a function $w \colon 2^E \to \mathbb{F}[X]$ by setting $w(E') := \prod_{e \in E'} w(e)$ for subsets $E'$ of $E$. The *generating function* $\mathrm{GF}(G, \mathscr{E})$ is defined as

$$\mathrm{GF}(G, \mathscr{E}) := \sum_{E' \subseteq E} w(E') \ ,$$

where the sum is over all subsets $E'$ of $E$ such that the spanning subgraph $(V, E')$ of $G$ has property $\mathscr{E}$.

Let $K_n$ denote the complete graph on $n$ nodes and the variable $X_{ji} = X_{ij}$ be the weight of the edge $\{i, j\}$ for $i \neq j$.

For most graph properties $\mathscr{E}$ of interest, the sequence $(\mathrm{GF}(K_n, \mathscr{E}))$ of generating functions is VNP-complete, see [Bür00a, Chap. 3]. For instance, let $\mathscr{DI}$ denote the graph property expressing that all connected components have exactly two nodes. The corresponding generating function enumerates the *perfect matchings*, or *dimer coverings*. More specifically, the generating function of perfect matchings in complete bipartite graphs is nothing but the permanent polynomial: $\mathrm{GF}(K_{n,n}, \mathscr{DI}) = \mathrm{PER}_n$.

Only in very few cases of interest $(\mathrm{GF}(K_n, \mathscr{E}))$ lies in VP. The graph property consisting of all trees is such an example. A well-known result due to Kirchhoff expresses the generating function as a determinant as follows. Consider the $n$ by $n$ matrix $A = [a_{ij}]$ by setting $a_{ij} := -X_{ij}$ for $i \neq j$ and $a_{ii} := -\sum_{\ell \neq i} X_{i\ell}$. Moreover, let $B$ be a matrix arising from $A$ by deleting one row and one column of the same index. Then we have (cf. [Bol79, p. 40])

$$\mathrm{GF}(K_n, \{\text{trees}\}) = \det(B).$$

This immediately implies that $(\mathrm{GF}(K_n, \{\text{trees}\})) \in \mathrm{VBP}$.

Another famous example is due to Fisher [Fis61] and Kasteleyn [Kas61]. It follows from the discovery that $\mathrm{GF}(G_n, \mathscr{DI})$ can be expressed as Pfaffian, see [Bür00a, §3.2].

**Theorem 2.34.** *The family* $(\mathrm{GF}(G_n, \mathscr{DI}))$ *is p-computable for any p-sequence* $(G_n)$ *of planar graphs.*

In fact, this result is the starting point of Valiant's theory of holographic algorithms [Val08], see Remark 2.32.

2.9. **Determinantal complexity.** The problem of deriving the permanent from the determinant by substitution is classical. First note that per $\begin{bmatrix} a & b \\ c & d \end{bmatrix} = \det \begin{bmatrix} a & -b \\ c & d \end{bmatrix}$. Pólya [Pól13] asked in 1913 whether such a formula is also possible for $n \geq 3$, i.e., whether there is a sign matrix $[\epsilon_{ij}]$ such that $\mathrm{per}_n = \det[\epsilon_{ij} x_{ij}]$. This was disproved by Szegő [Sze13] in the same year. Marcus and Minc [MM61] strengthened this result by showing that there is no matrix $[f_{pq}]$ of linear forms $f_{pq}$ in the variables $x_{ij}$ such that $\mathrm{per}_n = \det[f_{pq}]$.

**Definition 2.35.** The *determinantal complexity* $\mathrm{dc}(f)$ of a polynomial $f \in \mathbb{F}[x_1, \ldots, x_N]$ is the smallest $s$ such that there exists a square matrix $A$ of size $s$, whose entries are affine linear functions of $x_1, \ldots, x_N$, such that $f = \det(A)$. Moreover, we write $\mathrm{dc}(n) := \mathrm{dc}(\mathrm{PER}_n)$.

A priori, it is not even clear that $\mathrm{dc}(f)$ is finite. However, Proposition 2.23 shows $\mathrm{dc}(f) \leq E(f) + 1$. In combination with Ryser's formula (2.3), we obtain that $\mathrm{dc}(n) = O(n^2 2^n)$. Grenet [Gre12] showed that $\mathrm{dc}(n) \leq 2^n - 1$ by a nice direct combinatorial construction. This implies $\mathrm{dc}(3) \leq 7$, which was shown to be optimal in [IH16, ABV17]. It is easy to see that $dc(2) = 2$. Landsberg and Ressayre [LR17] recently proved that the representation $per_m = \det(A)$ in Grenet's construction is optimal among all representations respecting "half of the symmetries" of $per_m$.

We can equivalently rephrase Corollary 2.33 as follows (assuming char $\mathbb{F} \neq 2$): VBP $\neq$ VNP is true iff $dc(n)$ is not polynomially bounded in $n$.

Unfortunately, little is known with regard to lower bounding $dc(n)$. The following result due to Mignon and Ressayre [MR04] is the best known lower bound for $dc(n)$, except for a small improvement over $\mathbb{F} = \mathbb{R}$ due to Yabe [Yab15].

**Theorem 2.36.** *We have $dc(n) \geq n^2/2$ over $\mathbb{F} = \mathbb{C}$.*

The idea of the proof is to consider the rank of the Hessian of a polynomial. This also works over any field with char $\mathbb{F} \neq 2$ with some technical twist [CCL10]. The paper [LMR13] extended this lower bound to the border version of determinantal complexity, which required substantial new ideas.

## 3. Properties of complexity classes

We show here that the previously defined complexity classes are closed with respect to various operations. Moreover, we briefly discuss structural questions and the surprising fact that, over finite fields, there are natural problems of intermediate complexity.

### 3.1. **Robustness.**
It is easy to see that all the complexity classes introduced before are closed under $p$-projections and with respect to several natural operations, such as taking sums, products, or compositions. The latter means forming $(f_n(g_1, \ldots, g_{v(n)}))$ from sequences $(f_n)$ and $(g_n)$, where $f_n \in \mathbb{F}[x_1, \ldots, x_{v(n)}]$.

It is important that the class VNP is closed under taking coefficients [Val82]. The precise meaning of this is as follows. Consider a sequence of polynomials $f_n \in \mathbb{F}[X_n]$ in the set of variables $X_n$. For each $n$, fix a monomial formed from a subset $Y_n$ of $X_n$ and let $h_n$ denote the coefficient of this monomial in $f_n$. Thus $h_n$ is a polynomial in the remaining variables $X_n \setminus Y_n$. We call $(h_n)$ a coefficient sequence of $(f_n)$.

**Proposition 3.1.** *If $(f_n)$ is $p$-definable, then any of its coefficient sequences is $p$-definable as well.*

However, none of the classes VF, VBP and VP shares this property, if VP $\neq$ VNP. This can be seen from the following example. Consider

$$f_n := \prod_{i=1}^{n}\left(\sum_{j=1}^{n} x_{ij} y_j\right).$$

The family $(f_n)$ lies in VF since $E(f_n) = O(n^2)$. On the other hand, the coefficient of the product $Y_1 \cdots Y_n$ in $f_n$ equals the permanent $PER_n$.

Valiant also proved in [Val82] that VNP is closed under $p$-bounded applications of differentiation and integration. Again, the classes VF, VBP, and VP fail to be closed under these operations, as the following formula shows:

$$\frac{\partial}{\partial y_1} \cdots \frac{\partial}{\partial y_n} f_n = \left(\frac{3}{2}\right)^n \int_{-1}^{1} \cdots \int_{-1}^{1} y_1 \cdots y_n f_n \, dy_1 \cdots dy_n = PER_n.$$

### 3.2. **Complexity of factors.**
We focus now on the following general, but vaguely stated question: Suppose $f$ is a structured polynomial and $g$ is a factor of $f$. Does $g$ inherit some of the structure of $f$?

Surprisingly, if we interpret "structured" as meaning that $f$ is computable by small circuits, then the answer is yes. This is the outcome of a sequence of seminal works by Kaltofen [Kal85b, Kal87a, Kal87b, Kal89]. In fact, Kaltofen [Kal85b] had designed a probabilistic reduction of the problem of factoring multivariate polynomials given by straight-line programs to the bivariate factorization problem. The focus on the existence of small ciruits (ignoring the cost of their construction) leads to a proof considerably simpler than in Kaltofen's paper, see [Bür00a, Thm. 2.21].

**Theorem 3.2.** *Suppose the multivariate polynomial $f$ of degree $d$ over a field $\mathbb{F}$ of characteristic zero is computed by an arithmetic circuit of size $s$. Then any factor $g$ of $f$ can be computed by an arithmetic circuit of size polynomially bounded in $s$ and $d$.*

The key idea is to carry out a Hensel lifting, which can be interpreted as a homotopy continuation method with respect to a discrete valuation. (Homotopy continuation is a standard method in numerical analysis for solving nonlinear systems of equations, see for instance [AG90, BC13].)

**Corollary 3.3.** *The class VP over a field of characteristic zero is closed under taking factors: if $(f_n) \in$ VP and $g_n$ divides $f_n$ for all $n$, then $(g_n) \in$ VP.*

This closedness result is a crucial ingredient in the seminal hardness versus randomness results of Kabanets and Impagliazzo [KI04], see Section 6.

There is a recent line of works [DSY10, Oli16, CKS19, DSS22] concerned with extending Theorem 3.2 to formulas, branching programs or bounded-depth circuits. It remains unknown whether the classes VF and VBP are closed under taking factors. However, the following was shown recently [CKS19], confirming a conjecture by Bürgisser in [Bür00a].

**Theorem 3.4.** *The class VNP over a field of characteristic zero is closed under taking factors.*

Lipton and Stockmeyer [LS78] discovered that there exist polynomials $f$ having factors with a complexity exponential in the complexity of $f$. An example is $f_n = x^{2^n} - 1 = \prod_{j < 2^n}(x - \zeta^j)$, where $\zeta = \exp(2\pi i / 2^n)$. By repeated squaring we get $L(f_n) \leq n + 1$. On the other hand, one can prove that, for almost all $M \subseteq \{0, 1, \ldots, 2^n - 1\}$, $\prod_{j \in M}(x - \zeta^j)$ has complexity exponential in $n$. Therefore, the dependence on the degree $d$ in Theorem 3.2 is necessary. Bürgisser's Factor Conjecture [Bür00a, Conj. 8.3] claims that in this setting, any factor $g$ of $f$ can be computed by an arithmetic circuit of size polynomially bounded in $s$ and the degree of $g$. This conjecture naturally arises when trying to link decisional and computational problems in algebraic settings [Bür01]. Bürgisser proved in [Bür04] that the Factor Conjecture is true for the notion of approximate complexity, see Theorem 4.22. This result is relevant for hardness versus randomness results in geometric complexity theory [Mul17].

### 3.3. Families of intermediate complexity.

Ladner's classical theorem [Lad75] states the existence of problems of intermediate problems, assuming P $\neq$ NP. However, these problems are artificial and constructed by diagonalization. It is not surprising that similar results can be proven in the algebraic setting.

Definition 2.15 of the $p$-reduction is very restrictive since it only allows substitution. Instead, one may study an oracle based reduction in the style of Cook reductions. Let $(f_n)$ and $(g_m)$ be $p$-families of multivariate polynomials over the fixed field $\mathbb{F}$. We say that a $(f_n)$ is a *c-reduction* of $(g_m)$ iff there is a polynomially bounded function $t \colon \mathbb{N} \to \mathbb{N}$ such, for each $n$, there is an arithmetic circuit of polynomial size in $n$, that computes $f_n$ from $g_{t(n)}$, variables and constants. We call two $p$-families $(f_n)$ and $g_m)$ *c-equivalent* iff they are *c*-reductions of each other. The *c-degrees*, defined as *c*-equivalence classes of $p$-definable families, form a partially ordered set with respect to *c*-reduction. Valiant's conjecture VP $\neq$ VNP just expresses that there is more than one *c*-degree. If this the case, then any countable poset can be embedded in the poset of *c*-degrees [Bür99, Bür00a]. While the proof is a general diagonalization argument, the nonuniformity of the situation at hand allows for a short and quite elegant proof.

Surprisingly, over finite fields, one can say much more and exhibit *natural problems of intermediate difficulty*. Consider the *cut enumerator over* $\mathbb{F}_q$

$$\text{Cut}_n^q := \sum_S \prod_{i \in A, j \in B} x_{ij}^{q-1},$$

where the sum is over all cuts $S = \{A, B\}$ of the complete graph $K_n$ on the vertex set $\{1, \ldots, n\}$. To justify the naming, suppose an edge $\{i, j\}$ of $K_n$ carries the weight $w_{ij} \in \mathbb{N}$, define the weight of $S$ as the sum of the weights of all edges separated by $S$, and let $c(s)$ denotes the number of cuts with weight $s$. Then, under the substitution $x_{ij} \mapsto t^{w_{ij}}$, the cut polynomial becomes $\text{Cut}_n^q(t) := \sum_s (c(s) \bmod p) \, T^{(q-1)s}$, where $p$ is the characteristic of $\mathbb{F}_q$.

**Theorem 3.5.** *The family* $(\text{Cut}_n^q)_n$ *of cut enumerators is neither in* $\text{VP}^{\mathbb{F}_q}$ *nor* $\text{VNP}^{\mathbb{F}_q}$-*complete, provided* $\text{Mod}_p\text{NP} \nsubseteq \text{P/poly}$.

Here the class $\text{Mod}_p\text{NP}$ denotes the set of languages $\{x \in \{0,1\}^* \mid \phi(x) \equiv 1 \bmod p\}$, where $\phi$ is in the class $\#\text{P}$.

The idea behind the proof is that $\text{Cut}_n^q(t)$ can be evaluated over $\mathbb{F}_q$ by Boolean circuits of polynomial size. This relies on $\lambda^{q-1} = 1$ for all nonzero $\lambda \in \mathbb{F}_q$ (Fermat's little theorem). Note that this does not imply that $\text{Cut}_n^q(t)$ can also be evaluated by arithmetic circuits of polynomials size.

Theorem 3.5 was generalized in [MS18] to families related to vertex cover, clique, 3D matching etc.

## 4. Second generation algebraic complexity classes

Here we present modifications and extensions of the previously defined complexity classes. We begin with Malod's constant-free classes that are defined over $\mathbb{Z}$ and are most fundamental. Then we discuss classes that capture also polynomial families of exponentially growing degree (like resultants of systems of polynomials), again relying on Malod's work. We move on to discuss a complexity class conceived by Koiran and Poizat, which captures an analogue of PSPACE. Then we explain the known logical implication between the classical P $\neq$ NP conjecture, the one by Valiant, and the analogous conjectures in the Blum-Shub-Smale model. We proceed by discussing two conjectures that seem completely unrelated to complexity. Still both would imply the desired separation VP $\neq$ VNP. The first one has a number theoretic flavour and second one bounds the number of real zeros of real univariate polynomials given by depth-four circuits. Finally, we introduce the concept of the closure of a complexity class, which naturally enters in geometric complexity theory.

4.1. **Constant-free classes.** For comparing algebraic complexity classes over different fields and for relating them to other models of computation, it is important to work with a modification of the classes. These constant-free complexity classes were introduced and studied by Malod in his thesis [Mal03]. A *constant-free arithmetic circuit* $\Phi$ is a circuit as in Definition 2.1, but with the additional requirement that input gates carry labels from $X \cup \{-1, 0, 1\}$. That is, $-1, 0, 1$ are the only constants available. Clearly, the polynomials $\tilde{\Phi}_v$ computed by the circuit have integer coefficients. Let us define the *weight* $\text{wt}(f)$ of an integer polynomial $f$ as the sum of the absolute values of its coefficients. It is easy to check that the weight is subadditive and submultiplicative. The size of multiplicatively disjoint circuits controls the degree and weight: by induction, one easily shows:

**Lemma 4.1.** *We have* $\deg(\tilde{\Phi}_v) \leq |\Phi_v| + 1$ *and* $\log_2 \text{wt}(\tilde{\Phi}_v) \leq |\Phi_v| + \deg(v)$ *for all gates* $v$ *of a multiplicatively disjoint constant-free arithmetic circuit* $\Phi$.

**Definition 4.2.** The constant-free complexity classes $\text{VF}^0$, $\text{VBP}^0$, $\text{VP}^0$, and $\text{VNP}^0$ are defined as in Definitions 2.9 and 2.25, by additionally requiring the circuits $\Phi_n$ to be constant-free.

We note that these constant-free classes consist of sequences $(f_n)$ of multivariate polynomials with integer coefficients whose degree and bitsize grow at most polynomially in $n$. These classes are universal in the following sense: let $\mathbb{F}$ be a field. Then the sequences in the classes $\mathrm{VF}^{\mathbb{F}}$, $\mathrm{VBP}^{\mathbb{F}}$, $\mathrm{VP}^{\mathbb{F}}$, $\mathrm{VNP}^{\mathbb{F}}$ over $\mathbb{F}$ are obtained from sequences $(f_n)$ in the respective classes $\mathrm{VF}^0$, $\mathrm{VBP}^0$, $\mathrm{VP}^0$, $\mathrm{VNP}^0$ by applying to $f_n$ ring morphisms $\mathbb{Z}[x_1, \ldots, x_n, y_1, \ldots, y_m] \to \mathbb{F}[x_1, \ldots, x_n]$, which fix certain variables $x_i$ and map other variables $y_j$ to elements of $\mathbb{F}$.

Malod [Mal03] proved that $\mathrm{VNP}^0$ is closed under taking coefficients, that is, the analogue of Proposition 3.1 holds. We also note that Valiant's Criterion (Proposition 2.27) actually shows containement in $\mathrm{VNP}^0$.

Constant-free projections are defined by requiring $a_i \in \{-1, 0, 1\} \cup X$ in Definition 2.15 and the notion of $\mathrm{VNP}^0$-completeness is defined with respect to constant-free $p$-projections. The following was shown in [Mal03].

**Theorem 4.3.** *The family HC is $\mathrm{VNP}^0$-complete. Thus $\mathrm{VP}^0 = \mathrm{VNP}^0$ iff* $(\mathrm{HC}_n) \in \mathrm{VP}^0$.

This implies that HC is $\mathrm{VNP}^{\mathbb{F}}$-complete over any field $\mathbb{F}$. Indeed, the following general principle holds: if $(f_n)$ is $\mathrm{VNP}^0$-complete, then, over any field $\mathbb{F}$, $(f_n)$ defines a $\mathrm{VNP}^{\mathbb{F}}$-complete family. Note that we do not expect to prove that $(\mathrm{PER}_n)$ is $\mathrm{VNP}^0$-complete, since this would imply $\mathrm{VP}^{\mathbb{F}_2} = \mathrm{VNP}^{\mathbb{F}_2}$ (determinant and permanent coincide over $\mathbb{F}_2$).

### 4.2. Non-bounded degree: the classes $\mathrm{VP}_{\mathrm{nb}}^0$ and $\mathrm{VNP}_{\mathrm{nb}}^0$.

The sequences $(f_n)$ in all the classes defined so far have the property that the degree of $f_n$ is polynomially bounded in $n$. On the other hand, there are interesting and natural sequences of polynomials of exponential degree; here is an example.

**Example 4.1.** The *resultant* $R_n$ of a system of $n$ quadratic forms $f_i$ in $n$ variables is an irreducible integer polynomial in the $O(n^2)$ many coefficients of the system, which is homogeneous of degree $2^{n-1}$ in the coefficients of $f_i$ for each $i$. If the coefficients of the system are specialized to elements in an algebraically closed field $\mathbb{F}$, then the vanishing of this resultant is a necessary and sufficient condition for the existence of a nontrivial solution of the corresponding system of equations $f_1 = 0, \ldots, f_n = 0$ over $\mathbb{F}$. (Compare van der Waerden [Wae48, §82] or Lang [Lan93, Chap. IX].) Thus deciding whether the given system is feasible amounts to test whether the resultant $R_n$ evaluates to zero. See [GKP13] for complexity results.

If we remove in Definition 2.9 the assumption of the multiplicative disjointness, the resulting complexity class is denoted $\mathrm{VP}_{\mathrm{nb}}^0$ and was first defined and investigated by Malod [Mal03, Mal07].

**Definition 4.4.** The complexity class $\mathrm{VP}_{\mathrm{nb}}^0$ is defined as the set of sequences $(f_n)_{n \in \mathbb{N}}$ of multivariate polynomials such that there exists a sequence $(\Phi_n)_{n \in \mathbb{N}}$ of constant-free arithmetic circuits such that $\Phi_n$ computes $f_n$ and the size $|\Phi_n|$ is bounded by a polynomial in $n$. The complexity class $\mathrm{VNP}_{\mathrm{nb}}^0$ is defined via exponential summation as in Definition 2.25, requiring only $(g_n) \in \mathrm{VP}_{\mathrm{nb}}^0$.

The sequences $(f_n)$ in the classes $\mathrm{VP}_{\mathrm{nb}}^0$ and $\mathrm{VNP}_{\mathrm{nb}}^0$ consist of integer polynomials and the subscript "nb" stands for non-bounded degree. An important difference to $\mathrm{VP}^0$, $\mathrm{VNP}^0$ is that the degree and the bit size of $f_n$ may be exponentially growing in $n$. For instance, the sequence $(X+1)^{2^n} = \sum_i \binom{2^n}{i} X^i$ is in $\mathrm{VP}_{\mathrm{nb}}^0$.

We define the unbounded degree classes $\mathrm{VP}_{\mathrm{nb}}^{\mathbb{F}}$, $\mathrm{VNP}_{\mathrm{nb}}^{\mathbb{F}}$ over a field $\mathbb{F}$ by allowing circuits using constants in $\mathbb{F}$ for free. As for the bounded degree classes, these classes are obtained from the universal classes $\mathrm{VP}_{\mathrm{nb}}^0$, $\mathrm{VNP}_{\mathrm{nb}}^0$ by specializing certain variables to constants in $\mathbb{F}$. If $p = 2$, these classes have been studied in the context of Boolean functions.

*Remark* 4.5. $\mathrm{VP}_{\mathrm{nb}}^{\mathbb{F}_2}$ is the nonuniform version P/poly of polynomial time. Moreover, $\mathrm{VNP}_{\mathrm{nb}}^{\mathbb{F}_2}$ is the nonuniform version $\oplus$P/poly of parity polynomial time.

A straightforward way to obtain a sequence in $\mathrm{VP}_{\mathrm{nb}}^{\mathbb{F}}$ is to take a sequence $(f_n)$ in $\mathrm{VP}^{\mathbb{F}}$ and to apply substitutions $Z \mapsto X^{2^{\ell}}$ to the variables $Z$ of $f_n$, where $\ell$ is polynomially bounded in $n$ (repeated squaring). Similarly, applying such substitutions to a sequence in $\mathrm{VNP}^{\mathbb{F}}$, we obtain a sequence in $\mathrm{VNP}_{\mathrm{nb}}^{\mathbb{F}}$. One may ask whether all sequences in $\mathrm{VNP}_{\mathrm{nb}}^{\mathbb{F}}$ may be obtained this way. Corollary 4.7 below shows that this is indeed that case over finite fields!

While the definition of $\mathrm{VP}_{\mathrm{nb}}^0$ appears natural, the role of the class $\mathrm{VNP}_{\mathrm{nb}}^0$ is less clear. An essential feature justifying VNP is that it is closed under taking coefficients (Proposition 3.1). It is unknown whether $\mathrm{VNP}_{\mathrm{nb}}^0$ has this closure property, even though Malod [Mal07] showed that this property holds for the class $\mathrm{VNP}_{\mathrm{nb}}^{\mathbb{F}_p}$ arising when considering computations in a finite field $\mathbb{F}_p$.

**Theorem 4.6.** *The class $\mathrm{VNP}_{\mathrm{nb}}^{\mathbb{F}_p}$ is closed under taking coefficients, where $p$ is a prime.*

*Proof outline of Theorem 4.6, see* [Mal07]. For fixed $n$ we consider the generic computation $G_{-n} = 1, G_{-n+1} = x_1, \ldots, G_0 = x_n, G_1, \ldots, G_n$ defined recursively by

$$(4.1) \qquad G_m := \left( \sum_{i=-n}^{m-1} a_{mi} G_i \right) \left( \sum_{j=-n}^{m-1} b_{mj} G_j \right), \quad \text{for } 1 \leq m \leq n.$$

Here, the $x_1, \ldots, x_n$ are considered the input variables and the $a_{mi}, b_{mj}$ are auxiliary new variables (see [Bür00a, §5.6]). Note that $G_m$ can be computed by a constant-free circuit of size $O(mn + m^2)$. By specializing the auxiliary variables, it is easy to see that $(G_n)$ is a $\mathrm{VP}_{\mathrm{nb}}^0$-complete sequence. We derive from this the $\mathrm{VNP}_{\mathrm{nb}}^0$-complete sequence $(D_n)$ defined by

$$(4.2) \qquad D_n := \sum_{k=0}^{n} c_k \sum_{e \in \{0,1\}^{n-k}} G_n(x_1, \ldots, x_k, e_1, \ldots, e_{n-k}),$$

where the $c_i$ are new variables.

The key issue is to find a combinatorial description of the coefficients of the monomials in $G_n$. For expanding $G_m$ in (4.1), we have to select a left term $a_{mi} G_i$ and right term $b_{mj} G_j$, where $i < m$ and $j < m$. For further expanding the left copy of $G_i$, we have to select a right term and a left term in it, and similarly for the right copy of $G_j$. The process stops after expanding the various copies of $G_1$. The selection procedure can be modelled by a complete binary tree of depth $n$. More compactly, we can encode it by a digraph with $2n + 1$ nodes which we label by $-n, -n+1, \ldots, 0, 1, \ldots, n$. (Here, node $-n + i$ stands input $x_i$.) We put a "down" arrow pointing from $m$ to $i$ and an "up" arrow pointing from $m$ to $j$ when selecting $a_{mi} G_i$ in the left parenthesis of $G_m$, and $b_{mj} G_j$ in its right parenthesis. The expansion produces a copy of the monomial $x_1^{d_1} \cdots x_n^{d_n}$ in $G_n$, where $d_i$ is the indegree of the node labeled $-n + i$ (see Figure 4).

We denote by $A_{mi}$ the number of down arrows pointing from $m$ to $i$, and by $B_{mj}$ the number of up arrows pointing from $m$ to $j$, where $1 \leq m \leq n$ and $-n \leq i, j < n$. This defines two matrices $A, B$ of format $n \times 2n$ with nonnegative integer entries satisfying $A_{mi} = B_{mi} = 0$ if $i \geq m$. If we write $r_m(A) := \sum_i A_{mi}$ and $c_j(A) := \sum_m B_{mj}$ for the row and column sum of $A$, we check that

$$(4.3) \quad d_m = r_m(A) = r_m(B) = c_{-n+m}(A) + c_{-n+m}(B) \text{ for } 1 \leq m < n \text{ and } d_1 = r_m(A) = r_m(B) = 1.$$

Let us call pairs $(A, B)$ of matrices *valid* if they satisfy these constraints.

If we denote by $\mathrm{Multi}(\alpha_1, \ldots, \alpha_s) := (\sum_i \alpha_i)! / \prod_{i=1}^s \alpha_i!$ the multinomial coefficient, then the number of digraphs leading to a fixed pair $(A, B)$ of such matrices is given by the product of the multinomial coefficients of all the row distributions of $A$ and $B$:

$$(4.4) \qquad \prod_{m=1}^{n} \mathrm{Multi}(A_{m\bullet}) \, \mathrm{Multi}(B_{m\bullet}).$$

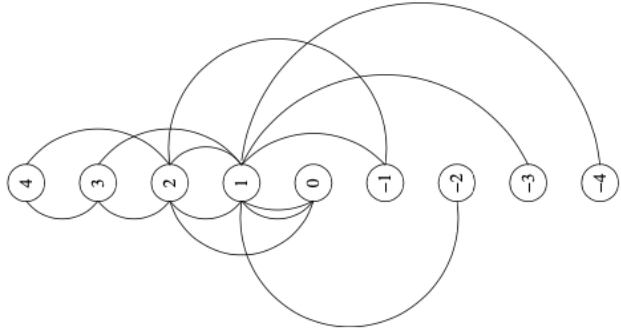

FIGURE 4. The digraph encoding the monomial $x_1 x_2 x_3^2 x_4^3$.

Summing these over all valid pairs $(A, B)$ of matrices is the coefficient of $x_1^{d_1} \cdots x_n^{d_n}$ in $G_n^n$ we are looking for.

The problem is now that the above product of multinomial coefficients can have exponential bitsize. since the degrees $d_m$ have bitsize as large as polynomially bounded in $n$. However, this problem disappears when we compute modulo a fixed prime $p$. A well known theorem by Lucas allows the efficient computation of binomial coefficients as follows. If we expand positive integers $\alpha, \beta$ with respect to the base $p$, $\alpha = \sum_s \alpha_s p^s$, $\beta = \sum_s \beta_s p^s$, then

$$\binom{\alpha}{\beta} = \prod_s \binom{\alpha_s}{\beta_s}.$$

This shows that the computation of a given monomial in $G_n^n$ can be carried out in #P/poly. The argument extends to coefficients of monomials of the $\text{VNP}^0$-complete sequence $(D_n)$.

Let us now expand the degrees $d_m = \sum_\ell d_{m\ell} 2^\ell$ in binary ($d_{m\ell} \in \{0, 1\}$). We can reduce the degrees by replacing the monomial $\prod_m x_m^{d_m}$ by

$$\prod_m \prod_\ell z_{m\ell}^{d_{m\ell}},$$

which has degree at most one in each variable $z_{m\ell}$. Note that the original monomial is obtained from this by substituting $z_{m\ell} \mapsto x_m^{2^\ell}$.

The assertion of the theorem now easily follows with Proposition 2.27.                              □

As a consequence of the proof, we obtain for sequences in $\text{VNP}_{\text{nb}}^{\mathbb{F}_p}$ a remarkable normal form computation via a sequence in $\text{VNP}^{\mathbb{F}_p}$.

**Corollary 4.7.** *Every sequence in* $\text{VNP}_{\text{nb}}^{\mathbb{F}_p}$ *is obtained from a sequence* $(f_n) \in \text{VNP}^{\mathbb{F}_p}$, *by applying substitutions of the form* $z \mapsto x^{2^\ell}$ *to the different variables* $Z$ *of* $f_n$, *where* $\ell$ *depends on* $Z$ *and is bounded by a polynomial in* $n$.

Of course, the major question is whether the separation $\text{VP}_{\text{nb}} \neq \text{VNP}_{\text{nb}}$ holds. In positive characteristic $p$, Malod [Mal07] proved that this is equivalent to Valiant's conjecture over $\mathbb{F}_p$.

**Theorem 4.8.** *In positive characteristic* $p$ *we have,*

$$\text{VP}^{\mathbb{F}_p} = \text{VNP}^{\mathbb{F}_p} \iff \text{VP}_{\text{nb}}^{\mathbb{F}_p} = \text{VNP}_{\text{nb}}^{\mathbb{F}_p}.$$

*Proof.* "⇒": This is immediate from Corollary 4.7.

"⇐": Let $(f_n) \in \text{VNP}^{\mathbb{F}_p}$. Then $(f_n) \in \text{VNP}_{\text{nb}}^{\mathbb{F}_p}$ and hence, by assumption, we have $(f_n) \in \text{VP}_{\text{nb}}^{\mathbb{F}_p}$. By homogenization (Proposition 2.2), we see that $(f_n) \in \text{VP}^{\mathbb{F}_p}$. □

*Remark* 4.9. By the homogenization trick, the implication $\text{VP}^{\mathbb{F}} = \text{VNP}^{\mathbb{F}} \Leftarrow \text{VP}_{\text{nb}}^{\mathbb{F}} = \text{VNP}_{\text{nb}}^{\mathbb{F}}$ holds over any field $\mathbb{F}$. Surprisingly, this implication is unclear for the constant-free classes. The homogenization trick breaks down since we cannot control the arising potentially huge coefficients! (E.g., in the model of computations over $\mathbb{Q}$, the bitsize of numbers does not matter.)

When trying to extend Theorem 4.6 to $\text{VP}_{\text{nb}}^0$, one faces the difficulty of coping with binomial (or multinomial) coefficients $\binom{n}{m}$, which have bitsize exponential in the bitsize of the integers $m, n$. As far as conditional statements are concerned, this difficulty can be overcome using a result from [Bür09]. We obtain the following somewhat subtle extension of Theorem 4.8, compare [KP11, Poi14]; the assertion (2) appears to be new.

**Theorem 4.10.** (1) *We have for the constant-free classes,*

$$\text{VP}^0 = \text{VNP}^0 \Rightarrow \text{VP}_{\text{nb}}^0 = \text{VNP}_{\text{nb}}^0.$$

(2) *Over any field $\mathbb{F}$ of characteristic zero, assuming the Generalized Riemann Hypothesis,*

$$\text{VP}^{\mathbb{F}} = \text{VNP}^{\mathbb{F}} \Longleftrightarrow \text{VP}_{\text{nb}}^{\mathbb{F}} = \text{VNP}_{\text{nb}}^{\mathbb{F}}.$$

*Proof outline.* 1. Consider the problem of computing the $i$th bit of $\binom{n}{m}$, where $i$ is given in binary expansion (which has only polynomial length). This can be solved in polynomial time with oracle calls to the counting hierarchy by [Bür09] (previously, the best known upper bound was PSPACE). If we assume $\text{VP}^0 = \text{VNP}^0$, then $\#\text{P} \subseteq \text{P}/\text{poly}$ and the counting hierarchy collapses to $\text{P}/\text{poly}$. We can now argue as in the proof of Theorem 4.6 to show that the computation of a monomial of $G_n^n$ can be done in $\text{P}/\text{poly}$. Using $\text{VP}^0 = \text{VNP}^0$, it follows as in the proof of Corollary 4.7 that the $\text{VNP}^0$-complete sequence $(D_n)$ can be obtained from a sequence in $\text{VP}^0$ by variable substitutions $z \mapsto x^{2^\ell}$ with polynomially bounded $\ell$. Hence $(D_n) \in \text{VP}_{\text{nb}}$.

2. $\text{VP}^{\mathbb{F}} = \text{VNP}^{\mathbb{F}}$ implies $\text{VP}^{\bar{\mathbb{F}}} = \text{VNP}^{\bar{\mathbb{F}}}$, where $\bar{\mathbb{F}}$ denotes the algebraic closure of $\mathbb{F}$. (This follows since HC is $\text{VNP}^{\mathbb{F}}$-complete for any field $\mathbb{F}$.) In [Bür00b] it was shown that this implies $\#\text{P} \subseteq \text{P}/\text{poly}$, under the Generalized Riemann Hypothesis, see also [Bür00a, Cor. 4.6]. As above, this implies the collapse of the counting hierarchy to $\text{P}/\text{poly}$. The rest of the argument is as for the first assertion. The direction "⇐" follows by Remark 4.9. □

4.3. **The class VPSPACE$^0$.** The resultant sequence $(R_n)$ of Example 4.1 is not known to lie in $\text{VNP}_{\text{nb}}^0$. This motivates the definition of a larger class VPSPACE, which mirrors PSPACE in the algebraic setting.

Let $f = \sum_a c(a)x_1^{a_1} \cdots x_n^{a_1}$ be a polynomial of degree at most $d$ and with integer coefficients bounded in absolute value by $2^d$. The *coefficient function* of $f$ takes an exponent vector $(a_1, \ldots, a_n)$ and an index $i$, all given in binary encoding (and thus of size $O(n \log d)$), and outputs the $i$th bit in the binary expansion of $c(a)$, with the convention that the sign of $c(a)$ is the 0th bit.

The following class was introduced by Poizat [Poi08] and by Koiran and Perifel [KP09b]. The following definition is from [KP09b].

**Definition 4.11.** The complexity class VPSPACE$^0$ is the set of sequences $(f_n)_{n \in \mathbb{N}}$ of multivariate integer polynomials such that there is a polynomial $p(n)$ satisfying:

(1) the number of variables of $f_n$ is bounded by $p(n)$,
(2) the degree and the bitsize of the coefficients of $f_n$ are bounded by $2^{p(n)}$,
(3) the coefficient function of $f_n$ is in PSPACE/poly.

The class $\mathrm{VPSPACE}^{\mathbb{F}}$ is obtained from $(f_n) \in \mathrm{VPSPACE}^0$ by substituting variables by elements of $\mathbb{F}$.

**Proposition 4.12.** *We have* $\mathrm{VNP}_{\mathrm{nb}}^0 \subseteq \mathrm{VPSPACE}^0$.

*Proof outline.* It suffices to show that the coefficient function of the $\mathrm{VNP}_{\mathrm{nb}}^0$-complete sequence $(D_n)$ defined in (4.2) is in PSPACE/poly. For this, it is sufficient to show this for the coefficient function of the $\mathrm{VP}_{\mathrm{nb}}^0$-complete sequence $(G_n^m)$, see (4.1). In the proof of Theorem 4.6, it was shown that the coefficient function $(G_n^m)$ can be written as an exponential summation over products of multinomials 4.4. We use now the fact that the computation of the $i$th bit of a binomial coefficient $\binom{m}{n}$ is possible in PSPACE. $\qquad\square$

The question of whether $\mathrm{VP}_{\mathrm{nb}}^{\mathbb{F}} = \mathrm{VPSPACE}^{\mathbb{F}}$ holds over a field $\mathbb{F}$, can be equivalently studied in the regime of sequences $(f_n)$ of polynomially growing degrees. If we denote by $\mathrm{VPSPACE}_b^{\mathbb{F}}$ the subclass of $\mathrm{VPSPACE}^{\mathbb{F}}$ obtained by requiring $\deg f_n$ to be polynomially bounded in $n$, then it is shown in [KP09b] that $\mathrm{VP}_{\mathrm{nb}}^{\mathbb{F}} = \mathrm{VPSPACE}^{\mathbb{F}}$ if and only if $\mathrm{VP}^{\mathbb{F}} = \mathrm{VPSPACE}_b^{\mathbb{F}}$.

A classical result in algebra states that the multivariate resultant $R_n$ of Example 4.1 can be expressed in terms of the determinants of Macaulay matrices [Mac94, Wae48]. (See [GKZ94, Cha93] for a modern view.) Using this, Canny [Can88] gave an algorithm to evaluate resultants that runs in PSPACE. With these methods, one can also show the following, see [GKP13].

**Theorem 4.13.** *The resultant sequence* $(R_n)$ *of Example 4.1 lies in* $\mathrm{VPSPACE}^0$.

*Remark* 4.14. An important work by Koiran [Koi96] implies that, under the Generalized Riemannian Hypothesis, the problem of testing the resultant $R_n$ for zero (for rational inputs), lies in the Arthur Merlin class AM. This class, which can be viewed as a randomized version of NP, is contained in the second level of the polynomial hierarchy. This indicates that $(R_n)$ is not expected to be complete in $\mathrm{VPSPACE}^0$.

*Remark* 4.15. Poizat [Poi08] gave an elegant characterization of $\mathrm{VPSPACE}^0$ in terms of sequences of polynomials computed by *summation circuits* of polynomial size. These are constant-free arithmetic circuits with summation gates: on input a polynomial $f$ such gates carry out the summation $f|_{x=0} + f|_{x=1}$, where $x$ is a variable occuring in $f$. Note that if allowing the use of such gates at the end of the computation only, the class $\mathrm{VNP}_{\mathrm{nb}}^0$ is obtained. Much earlier, Babai and Fortnow [BF91] used a similar model to give a unified approach to arithmetization of Boolean circuits.

The class $\mathrm{VPSPACE}^0$ can also be neatly characterized more algebraically in terms of circuits of polynomially bounded depths [KP09b].

**Proposition 4.16.** *The complexity class* $\mathrm{VPSPACE}^0$ *equals the set of sequences of multivariate integer polynomials computed by a sequence* $(\Phi_n)$ *of constant-free arithmetic circuits such that*

(1) *the number of input gates and the depth of* $\Phi_n$ *are bounded by a polynomial in* $n$,

(2) *the computation of* $\Phi_n$ *can be done in* PSPACE/poly.

We remark that, unlike for Definitions 2.9 and 4.2, it is necessary to put a restriction on the generation of the circuits $\Phi_n$.

4.4. **Connection to Blum-Shub-Smale model.** We already mentioned the BSS-model [BSS89, BCSS98] in the introduction. This centers around the definition of algebraic complexity classes $\mathrm{P}_{\mathbb{F}}$ and $\mathrm{NP}_{\mathbb{F}}$ over (possibly ordered) fields $\mathbb{F}$ and the separation hypotheses $\mathrm{P}_{\mathbb{F}} \neq \mathrm{NP}_{\mathbb{F}}$. For $\mathbb{F} = \mathbb{R}$ and $\mathbb{F} = \mathbb{C}$, these complexity classes are defined by a variant of real or complex random access machines, or alternatively (and more elegantly), by uniform families of algebraic circuits [Poi95]. We refer to the mentioned references for more details. The standard $\mathrm{NP}_{\mathbb{R}}$-complete problem is the feasibility problem for a system of real polynomial inequalities, while the standard $\mathrm{NP}_{\mathbb{C}}$-complete problem is the feasibility problem for a system of complex polynomial equations. For $\mathbb{F} = \mathbb{F}_2$, the usual complexity classes P and NP, defined via Turing machines, are obtained. Thus the BSS-complexity classes over $\mathbb{C}$ and $\mathbb{R}$ capture the complexity of basic decisional problems in complex and real algebraic geometry, respectively.

The notion of an *algebraic circuit* extends arithmetic circuits by including sign nodes of indegree one. The sign of $x \in \mathbb{R}$ equals 1 if $x \geq 0$ and 0 otherwise. By the sign of $z \in \mathbb{C}$ we understand 1 if $z \neq 0$ and 0 otherwise. An equivalent model of computation, distinguishing between real and Boolean data types, was described in [Gat86] and called arithmetic networks. Let us point out that Valiant's model is conceptually simpler in that it does only involve arithmetic computations but no sign tests.

There were several attempts to clarify the logical connections between the separation hypotheses in these different models, compare [Bür01]. It is not hard to see that NP $\not\subseteq$ BPP implies $P_\mathbb{C} \neq NP_\mathbb{C}$, see [ABKPM09]. This relies on the fact that computations of algebraic circuits on integers can be efficiently simulated on Boolean inputs, by computing modulo sufficiently large random primes (Schwartz-Zippel-Lemma). An additional technical difficulty, which can be overcome, is the efficient elimination of finitely many complex constants that could potentially be helpful. So a separation of classical complexity classes implies a separation of BSS-classes over $\mathbb{C}$. In a similar direction, Bürgisser [Bür00b] proved that NP $\not\subseteq$ BPP implies $VP_\mathbb{C} \neq VNP_\mathbb{C}$, under the generalized Riemann hypothesis. The latter hypothesis is used for the elimination of potentially helpful complex constants, similarly as it was done by Koiran [Koi96] in his analysis of the complexity of Hilbert's Nullstellensatz.

Interestingly, when focussing on classes related to PSPACE, one can relate separations in the BSS-model to separations in Valiant's model and vice versa. More specifically, one can define a complexity class $PAR_\mathbb{C}$ via families of algebraic circuits of polynomially bounded depth [BCSS98, §18.2, Def. 4]. Koiran and Perifel [KP09a] proved that the separation $P_\mathbb{C} \neq PAR_\mathbb{C}$ in the BSS-model implies the separation $VP_{nb}^0 \neq VPSPACE^0$ in Valiant's model (strictly speaking, for the uniform versions of these classes). Over $\mathbb{R}$, an analogous implication holds [KP09b]. The gist of these proofs is a way to replace the sign gates by oracle calls that evaluate the sign of polynomials from a fixed VPSPACE-family. So if we want to separate $P_\mathbb{C}$ from $PAR_\mathbb{C}$, we should first be able to separate the corresponding classes in Valiant's setting.

In the reverse direction, [Bür04, Cor. 4.5] proved that the separation $VNP^\mathbb{C} \not\subseteq \underline{VP}^\mathbb{C}$ in Valiant's model implies the separations $P_\mathbb{R} \neq PAR_\mathbb{R}$ as well as $P_\mathbb{C} \neq PAR_\mathbb{C}$ in the BSS-model. (For the border complexity class, see Definition 4.23 below.) The proof of the stated implication is a consequence of Theorem 4.22 on the border complexity of factors.

Summarizing, all these reasonings show that Valiant's algebraic model is closely tied to the BSS-model.

### 4.5. Counting classes for algebraic geometry.

Valiant's [Val79b] famous counting complexity class #P formalizes counting problems in combinatorics. In algebraic combinatorics, in the context of the representation theory of symmetric groups, certain nonnegative integers naturally as multiplicities. Famous examples are plethysms and Kronecker coefficients, which are of relevance in geometric complexity theory, see Section 7.4. Stanley [Sta00, §3] asked whether there are "positive formulas" for these quantities. A rigorous way to phrase this question is to ask whether the computation of these quantities is in the complexity class #P. It is not hard to show that plethysms and Kronecker coefficients can be expressed as the difference of two functions in #P, but is unclear whether they are in #P, see the recent research [IP22, IPP24]. We remark that the computation of plethysm and Kronecker coefficients lies in the quantum complexity class #BQP, see [BCG+24, IS24].

Counting problems are also important in algebraic geometry, where they describe the number of possible geometric configurations, see [EH16]. This prompts the question whether the computational complexity of such problems can be captured by appropriate complexity classes. This is indeed possible. Beginning with [Mee00, BC03], counting classes were introduced in the BSS-model and natural computational problems in complex and real algebraic geometry were identified as complete problems [BC06, BCL05]. Those are the computation of the degree and of the Euler characteristic of complex algebraic varieties, as well as the computation of the Borel-Moore Euler Characteristic of semialgebraic sets. Moreover, [BL07] shows that computing the Hilbert polynomial of smooth equidimensional projective varieties can be reduced in polynomial time to counting the number of complex zeros of systems of polynomials. These results are not limited to the BSS-model: the polynomial time reductions set up in the above papers also work in the Turing model of computation, leading to completeness results for the above mentioned problems in two complexity classes, whose precise relation to #P is still unclear, see [BC04, Problem 8.1].

4.6. **Tau Conjectures.** Shub and Smale [SS95] discovered a fascinating connection between a conjecture of a number theoretic flavour and the $P_{\mathbb{C}} \neq NP_{\mathbb{C}}$ conjecture. The *tau-complexity* $\tau(f)$ of an integer coefficient polynomial $f$ is defined as the minimal size of a constant-free arithmetic circuit computing $f$. (Recall that $-1, 0, 1$ are the only freely available constants in this model, compare Section 4.1.)

The $\tau$-*conjecture* claims the following connection between the number $z(f)$ of distinct integer roots of a nonzero univariate polynomial $f \in \mathbb{Z}[x]$ and its complexity $\tau(f)$:

$$(4.5) \qquad\qquad z(f) \leq (1 + \tau(f))^c,$$

for some universal constant $c > 0$ (compare also [Str90, Problem 9.2]). In [SS95] it was shown that the $\tau$-conjecture implies $P_{\mathbb{C}} \neq NP_{\mathbb{C}}$. In fact, the proof shows that in order to draw this conclusion, it suffices to prove that for all nonzero integers $m_n$, the sequence $(m_n n!)_{n \in \mathbb{N}}$ of multiples of the factorials is hard to compute. Hereby we say that a sequence $(a(n))$ of integers is *hard to compute* iff $\tau(a(n))$ is not polynomially bounded in $\log n$.

Motivated by [Koi04], Bürgisser [Bür09] proved the following result.

**Theorem 4.17.** *The $\tau$-conjecture implies* $VP^0 \neq VNP^0$. *Moreover,* $VP^0 \neq VNP^0$ *follows if the sequence $(n!)$ of factorials is hard to compute.*

As in [ABKPM09], the key idea for the proof is the consideration of the counting hierarchy, which was introduced by [Wag86]. This is a complexity class lying between P and PSPACE that bears more or less the same relationship to #P as the polynomial hierarchy bears to NP.

It is plausible that $(n!)$ is hard to compute, otherwise factoring integers could be done in (nonuniform) polynomial time, cf. Strassen [Str76] in [BCSS98, p.126]. Lipton [Lip94] strengthened this implication by showing that if factoring integers is "hard on average" (a common assumption in cryptography), then a somewhat weaker version of the $\tau$-conjecture follows.

The $\tau$-conjecture is false when replacing "integer zeros" by "real zeros". Koiran observed that when restricting to polynomials given by depth-four circuits, an analogous conjecture bounding real zeros has the same dramatic consequences. More specifically, we consider real univariate polynomials of the form

$$(4.6) \qquad\qquad F = \sum_{i=1}^{m} \prod_{j=1}^{k} f_{ij}$$

where all $f_{ij}$ are $t$-sparse, i.e., have at most $t$ monomials. Descartes rule states that a $t$-sparse polynomial $f$ has at most $t-1$ positive real zeros, no matter what is the degree of $f$. (This bound is optimal.) Therefore, a product $f_1 \cdots f_k$ of $k$ many $t$-sparse polynomials can have at most $k(t-1)$ positive real zeros. But what can we say about the number of real zeros of a sum of $m$ many such products?

The following conjecture was proposed by Koiran [Koi11].

**Conjecture 4.1** (Real $\tau$-conjecture). *The number of real zeros of a nonzero polynomial $F$ of the form (4.6) is upper bounded by a polynomial in $m$, $k$, and $t$.*

Koiran [Koi11] proved that the real $\tau$-conjecture implies the separation $\text{VP}^0 \neq \text{VNP}^0$ over $\mathbb{C}$; in fact, even $\text{VP} \neq \text{VNP}$ follows; see [Tav14]. Tavenas also observed that an upper bound polynomial in $m, t, 2^k$ is sufficient to derive this conclusion [Tav14, §2.1, Cor. 3.23]. For known upper bounds on the number of real zeros of polynomials of the form $F$, we refer to [KPT15] and the references given there. Hrubes [Hru13] identified statements equivalent to the real $\tau$-conjecture that are related to complex zero counting. In [BB20] it was shown that polynomials of the form (4.6) with independent standard Gaussian coefficients satisfy the claimed bound of Conjecture 4.1.

Recently, Dutta [Dut21] proposed a version of the real $\tau$-conjecture for linear combinations of squares of univariate polynomials $F \in \mathbb{R}[x]$. Consider representations

$$F = c_1 f_1^2 + \ldots + c_s f_s^2,$$

where $c_i \in \mathbb{R}$ and $f_i \in \mathbb{R}[x]$ has $t_i$ monomials. The *support-sum size* $\text{SoS}_{\mathbb{R}}(F)$ is defined as minimum of the sum of sparsities $t_1 + \ldots + t_s$ over all such representations of $F$. It is easy to see that a polynomial $F$ with $t$ monomials satisfies $\sqrt{t} \leq \text{SoS}_{\mathbb{R}}(F) \leq 2t + 2$. Here is Dutta's conjecture:

**Conjecture 4.2** (SoS $\tau$-conjecture)**.** *There is $c > 0$ such that for all nonzero $F \in \mathbb{R}[x]$, the number of real zeros of $F$ is at most $c \cdot \text{SoS}_{\mathbb{R}}(F)$.*

Remarkably, as shown in [Dut21], this conjecture not only implies the separation $\text{VP} \neq \text{VNP}$ over $\mathbb{C}$, but also implies the existence of explicit families of rigid matrices. Matrix rigidity was introduced by Valiant [Val77], as an auxiliary concept for proving lower complexity bounds for evaluating linear maps, most prominently the discrete Fourier transform. Roughly, a matrix is rigid, if it is far (in terms of Hamming distance) from any low rank matrix. Rigid matrices are easily shown to exist, yet their explicit construction is a major open question. In [Dut21] it is also shown that a variant of Conjecture 4.2 for sum of cubes implies the solution of another major open problem, namely that the polynomial identity testing problem (PIT) can be solved in deterministic polynomial time (compare Section 6).

4.7. **Closures of complexity classes.** *Border complexity* naturally enters the scene when proving lower complexity bounds with tools from algebraic geometry. In fact, almost all of the known techniques for proving complexity lower bounds actually provide more, namely lower bounds for border complexity. So it is not surprising that geometric complexity theory [MS01, MS08, Mul11] actually attempts to prove lower bounds on border complexity, see Section 7. In particular, *border rank*, the pendant of tensor rank, is a crucial concept in the development of asymptotically fast algorithms for matrix multiplications [BCLR79, Sch81, Str87a, CW90, Wil12, LG14]. See [BCS97, Chap. 15] and [Lan17] for further information. Originally, border complexity had been introduced in this context by Strassen [Str74] and was further studied in [Gri86, Lic90, Bür04].

In a nutshell, the notion of border rank arises naturally, when, besides arithmetic operations, one also allows limit processes with respect to a topology. Let us first rigorously define border rank in a topological way. The space of polynomials $P_{n,d} := \mathbb{F}[x_1, \ldots, x_n]_{\leq d}$ of degree at most $d$ is a finite dimensional $\mathbb{F}$-vector space and carries several topologies of interest. If $\mathbb{F} = \mathbb{C}$ there is the Euclidean topology. On the other hand, there is the Zariski topology defined over infinite fields[9]. Unfortunately, the complexity measures defined so far do not behave nicely with respect to these topologies. The reason is that the subsets $\{f \in P_{n,d} \mid L(f) \leq s\}$ are not closed. This not only applies to the complexity measure $L$ from Definition 2.4, but also to the corresponding notions obtained for formulas, weakly-skew circuits etc. This phenomenon was first observed in complexity theory when studying tensor rank in connection with the complexity of the matrix multiplication problem and led to the introduction of the notion of border rank, see [BCS97, Lan11, Blä13]. For instance, the trilinear form $f = x_1y_1z_2 + x_1y_2z_1 + x_2y_1z_1$ has rank three, but, $f$ can be approximated arbitrary closely by the forms $g_\varepsilon := \varepsilon^{-1}\big((x_1 + \varepsilon x_2)(y_1 + \varepsilon y_2)(z_1 + \varepsilon z_2) - x_1y_1z_1\big) = f + O(\varepsilon)$ of rank two, for $\varepsilon \to 0$.

---

[9]Over finite fields, the Zariski topology is uninteresting since any subset is open.

For these reasons, it is natural to introduce approximate notions of complexity.

**Definition 4.18.** The *border complexity* (also called *approximate complexity*) $\underline{L} \colon P_{n,d} \to \mathbb{N}$ is defined by the property that, for all $s$, $\{f \in P_{n,d} \mid \underline{L}(f) \le s\}$ is the closure of the set $\{f \in P_{n,d} \mid L(f) \le s\}$ for the Zariski topology.

This means that $\underline{L}(f) \le L(f)$ and that $\underline{L}(f)$ is the largest lower semicontinuous function of $f$ bounded by $L(f)$.

The following result is an essential starting point of geometric complexity theory.

**Theorem 4.19.** *Suppose $\mathbb{F} = \mathbb{C}$. If we take in Definition 4.18 the closures with respect to the Euclidean topology, we get the same notion $\underline{L}$.*

*Proof.* This follows from a general principle in algebraic geometry [Mum88, §10], which states that the Zariski closure of a constructible subset of $\mathbb{C}^N$ coincides with its Eulidean closure. $\qquad\square$

There is an algebraic way to characterize border complexity that we describe next. Let $\varepsilon$ denote a variable, $\mathbb{F}(\varepsilon)$ the field of rational functions in $\varepsilon$ over $\mathbb{F}$, and let $R \subseteq \mathbb{F}(\varepsilon)$ denote the local subring consisting of the rational functions $F$ defined at $\epsilon = 0$. For a polynomial $F$ with coefficients in $R$, the substitution $\epsilon \mapsto 0$ is a well defined homomorphism and yields the polynomial $F_{\epsilon=0}$ over $\mathbb{F}$. Instead of $\mathbb{F}(\varepsilon)$, one may equivalently consider the local ring $\mathbb{F}[[\varepsilon]]$ of formal power series in $\varepsilon$ over $\mathbb{F}$, which is often more convenient, see [Bür04, §5.2].

The following result is due to Alder [Ald84]; see also [BCS97, §20.6].

**Theorem 4.20.** *Let $f \in \mathbb{F}[x_1, \ldots, x_n]$. The border complexity $\underline{L}(f)$ of $f$ is the smallest natural number $s$ such that there exists $F$ in $R[x_1, \ldots, x_n]$ satisfying $F_{\epsilon=0} = f$ and $L(F) \le s$. Here the complexity $L$ is to be interpreted with respect to the larger field of constants $\mathbb{F}(\varepsilon)$.*

*Remark* 4.21. Passing to limits of polynomials leads to to concept of inital ideals. This is an important tool in commutative algebra and underlies the theory of Gröbner bases, see [Eis95]. Let us show that this is closely related to border complexity. The convex hull of the support $\operatorname{supp} f$ of a polynomial

$$f = \sum_{a \in \operatorname{supp} f} c(a) x_1^{a_1} \cdots x_n^{a_n} \qquad\qquad (c_a \neq 0)$$

is called the Newton polytope $P$ of $f$. To a supporting hyperplane $H$ of $P$ we may assign the corresponding initial term polynomial

$$\operatorname{in}_H f := \sum_{a \in H \cap \operatorname{supp} f} c(a) x_1^{a_1} \cdots x_n^{a_n}.$$

We claim that

$$\underline{L}(\operatorname{in}_H f) \le L(f) + n + 1.$$

Indeed, we may obtain $\operatorname{in}_H(f)$ as a "degeneration" of $f$ as follows. Assume that $\langle w, x \rangle - b = 0$ is the equation of $H$, say $\langle w, x \rangle \ge b$ on $P$. We can always achieve that $w \in \mathbb{Z}^n$, $b \in \mathbb{Z}$. Then we have

$$F := \varepsilon^{-b} f(\epsilon^{w_1} x_1, \ldots, \varepsilon^{w_n} x_n) = \sum_{a \in \operatorname{supp} f} c(a) \varepsilon^{\langle w, a \rangle - b} x_1^{a_1} \cdots x_n^{a_n} = \operatorname{in}_H f + O(\varepsilon)$$

using the convenient, intuitive Big-Oh notation. Therefore, $F_{\varepsilon=0} = \operatorname{in}_H f$ and $L(F) \le L(f) + n + 1$ which proves our claim. (Recall that the powers of $\varepsilon$ are considered as constants.)

The following result from [Bür04] on the complexity of factors shows the usefulness of border complexity when studying the complexity of factors of polynomials.

**Theorem 4.22.** *Suppose* $f, g \in \mathbb{F}[x_1, \ldots, x_n]$ *are such that* $g$ *divides* $f$ *and* $f \neq 0$. *Then* $\underline{L}(g)$ *is bounded by a polynomial in* $\underline{L}(f)$ *and* $\deg g$.

The motivation for this result was the *Factor Conjecture* [Bür00a, Conj. 8.3], which claims that Theorem 4.22 also holds for complexity. We refer to [DSS22] for a result in this direction.

Of course, new complexity measures call for the introduction of new complexity classes. The following class was first defined by Bürgisser [Bür04].

**Definition 4.23.** The complexity class $\underline{\mathrm{VP}}^{\mathbb{F}}$ is defined as the set of sequences $(f_n)_{n \in \mathbb{N}}$ of multivariate polynomials over $\mathbb{F}$ such that $\deg f_n$ and $\underline{L}(f_n)$ are polynomially bounded in $n$. The classes $\underline{\mathrm{VF}}^{\mathbb{F}}$ and $\underline{\mathrm{VBP}}^{\mathbb{F}}$ are defined in an analogous way.

It is an open question whether $\underline{\mathrm{VP}}$ is strictly larger than VP, see [GMQ16]. This is a formal way to ask whether the ability to pass to limits significantly increases computational power. In [Bür04, Bür20] it was observed that this question is closely related to a notion of approximation order. (If this order happened to be polynomially bounded, then the above classes would coincide.) The goal of geometric complexity [MS01, MS08, Mul11] is to prove the strengthening of Valiant's Conjecture: VNP $\not\subseteq \underline{\mathrm{VBP}}$. The related weaker conjecture VNP $\not\subseteq \underline{\mathrm{VP}}$ already appeared in [Bür01, Conj. 1]. We remark that $\underline{\mathrm{VP}}$ can be formally defined as the closure of VP with respect to the box topology [IS22].

The paper [BIZ18] contains remarkable characterizations of the class $\underline{\mathrm{VF}}$ in terms of complete problems, with a lovely link to Fibonacci sequences. The question on the need of approximations is currently intensely studied for restricted models of computation. For this, we refer to the recent papers [Kum20, BDI21, DDS22, DS22, DGI$^+$24a].

## 5. Restricted models of computation

By restricting the power of models of computation, one tries to isolate which aspects allow to increase efficiency and hopefully to prove complexity lower bounds in the restricted models. In the past decade, there has been intense research in restricted models of arithmetic circuits: due to lack of space, our discussion has to be brief. For more details we refer to the surveys by Shpilka and Yehudayoff [SY10] and Saptharishi [Sap15].

**5.1. Monotone circuits.** The efficiency of algebraic algorithms seems to rely on exploiting cancellations. In order to turn this observation into a rigorous statement, one defines an arithmetic circuit to be *monotone* if it only uses additions and multiplications, but no subtractions. Valiant [Val80] proved an exponential lower bound for monotone arithmetic circuits computing perfect matching polynomials for planar graphs, thus rigorously showing that an exponential gain is possible when allowing subtractions. The efficiency in the presence of subtractions results from the fact that the generating function of perfect matching in planar graphs can be expressed in terms of Pfaffians, see Theorem 2.34. Jerrum and Snir [JS82] even showed that the complexity of prominent functions like the permanent and iterated matrix multiplication can be *exactly* determined in the monotone model of computation. So monotone computations are too restrictive to be of interest for computational complexity.

**5.2. Multilinear circuits.** Permanents and determinants are both multilinear functions. What can we say if we require the arithmetic circuits to respect multilinearity? An arithmetic circuit $\Phi$ is called *syntactically multilinear* if for every product gate $v$ in $\Phi$ with sons $v_1$ and $v_2$, the set of variables $X_{v_1}$ and $X_{v_2}$ that occur in $\Phi_{v_1}$ and $\Phi_{v_2}$, respectively, are disjoint; see Definition 2.1. Clearly, such $\Phi$ is *semantically multilinear*, i.e., it computes multilinear polynomials at each gate. (For formulas this distinction is not relevant, however, this it is unclear for general circuits.)

In a breakthrough paper, Raz [Raz09] proved a superpolynomial lower bound for multilinear arithmetic formulas computing the permanent. Unfortunately, the same lower bound holds for the determinant. The proof combines the rank method for the matrix of partial derivatives with a surprising use of an unbalanced random partition of the set of variables. We note that tracing the rank of the matrix of partial derivatives was first introduced by Nisan [Nis91] for noncommutative formula size and later used by [NW97] for studying depth three circuits. In [RSY08] this method was applied to show a nonlinear lower bound for syntactically multilinear circuits computing an explicit polynomial and the proof was simplified in [RY09]. A refinement of these techniques led to the separation of the models of arithmetic formulas and arithmetic branching programs in the multilinear setting [DMPY12]: an explicit family $(f_n)$ of $n$-variate polynomials is constructed that has multilinear algebraic branching programs of size $O(n)$, but every multilinear formula computing $f_n$ must be of size $n^{\Omega(\log n)}$. However, we note that VP and VNP collapse in the very restricted world of multilinear computations [MST16].

5.3. **Set-multilinear circuits.** We now discuss set-multilinear polynomials and circuits. A map $[n]^d \to \mathbb{F}$, $(i_1, \ldots, i_d) \mapsto t_{i_1 \cdots i_d}$ defines a tensor $t$ of order $d$ over $\mathbb{F}$.[10] We can identify the tensor $t$ with the multilinear polynomial

$$f := \sum_{i_1, \ldots, i_d = 1}^{n} t_{i_1 \ldots i_d} X_{i_1}^{(1)} \cdots X_{i_d}^{(d)}$$

in the variables $X_j^{(i)}$. Note that the set of variables is naturally partitioned into $d$ classes such that each monomial of $f$ is a product of $d$ variables, each coming from exactly one of the classes. One calls $f$ *set-multilinear* with respect to this partition of variables. The *tensor rank $R(t)$* of $t$ is defined as the minimum $r$ such that there is an expression

$$(5.1) \qquad\qquad f = \sum_{\rho=1}^{r} u_\rho^{(1)} \cdots u_\rho^{(d)},$$

in which each $u_\rho^{(i)}$ is a linear form in the variables $X_1^{(i)}, \ldots, X_n^{(i)}$ of the $i$th class. Note that we may interpret (5.1) as a depth three formula of structure $\Sigma\Pi\Sigma$, which respects the set multilinearity.

Motivated by the problem to understand the complexity of matrix multiplication, tensors of order $d = 3$ have been a center of focus in algebraic complexity. Indeed, such tensors describe bilinear maps $\mathbb{F}^n \times \mathbb{F}^n \to \mathbb{F}^n$, for which matrix multiplication is an important example. In a seminal paper, Strassen [Str73b] proved that the arithmetic circuit complexity of $f$ is closely related to the tensor rank $R(t)$. Strassen's proof can be interpreted as a method of transforming any arithmetic circuit computing $f$ (which may even involve divisions) into a formula (5.1) of the above set-multilinear form.

Raz [Raz13] vastly generalized this method of "set multilinearization". Let $f$ be a multilinear polynomial over $d$ classes of variables. Then any formula of size $s$ and product-depth $\Delta$ computing $f$ can be transformed into a set-multilinear formula of size $O((\Delta+2)^d s)$, while preserving the product-depth $\Delta$. Using this, Raz established a connection between formula size and tensor rank: he showed that the existence of an explicit family $(t_n)$ of order $d_n$ tensors over $\mathbb{F}$ with $d_n \leq \log n / \log \log n$, such that $R(t_n) \geq n^{d_n(1-o(1))}$, would separate the classes VF and VNP.

5.4. **Noncommutative model.** When wondering about the gain provided by commutativity, one may study computations in the noncommutative polynomial ring over a field $\mathbb{F}$. (Its natural realizations are rings of matrices over $\mathbb{F}$.) In an influential paper, Nisan [Nis91] proved exact exponential lower bounds for the noncommutative formula size of the permanent and determinant (more generally, for noncommutative algebraic branching programs). This paper introduced the method of the rank of the matrix of partial derivatives, which has been applied in many situations. Let us point out that [LMS16] extends Nisan's result to a notion of skew circuits.

A famous paper by Strassen [Str73b] studies arithmetic circuits computing polynomials using divisions and how to eliminate them. In Hrubes and Wigderson [HW14] such questions are studied for noncommutative polynomials and their quotients. Remarkably, a characterization of noncommutative formula size in terms of the inverse of matrices holds, which is similar to Valiant's completeness result for the determinant, see Cohn [Coh85].

---

[10]This generalizes the concept of matrices obtained for $d = 2$.

In this context, Hrubes, Wigderson and Yehudayoff [HWY11] discovered a beautiful and surprising connection of noncommutative circuit size to the sum-of-squares problem, that we want to briefly describe. Given a field $\mathbb{F}$, we define $S_{\mathbb{F}}(n)$ as the smallest $m$ such that there exists a polynomial identity

$$(5.2) \qquad (x_1^2 + \ldots + x_n^2) \cdot (y_1^2 + \ldots + y_n^2) = z_1^2 + \ldots + z_m^2,$$

where each $z_k$ is a bilinear form in the variables $x_i$ and $y_j$, thus $z_k = \sum a_{ijk} x_i y_j$ with $a_{ijk} \in \mathbb{F}$. It is easy to see that $n \leq S_{\mathbb{F}}(n) \leq n^2$ if $\mathbb{F}$ has characteristic different from two. Suppose now $\mathbb{F} = \mathbb{R}$. The multiplicativity of the norm of complex numbers, $|a|^2 |b|^2 = |ab|^2$, translates to an identity showing $S_{\mathbb{R}}(2) \leq 2$. Similar identities for the quaternions and octonians express that $S_{\mathbb{R}}(4) \leq 4$ and $S_{\mathbb{R}}(8) \leq 8$. A famous result by Hurwitz [Hur22] states that, besides the trivial $k = 1$, these are all cases where $S_{\mathbb{R}}(k) \leq k$. Thus $S_{\mathbb{R}}(k) > k$ if $k \notin \{1, 2, 4, 8\}$. Recently, Hrubes [Hru24] proved that $S_{\mathbb{F}}(n) = O(n^{1.62})$ for $\mathbb{F} = \mathbb{C}$ and fields of positive characteristic.

More generally, we define $S_{\mathbb{F}}(f)$ for a biquadratic polynomial $f$ as the minimal $m$ such that $f$ can be written as a sum of squares $f = z_1^2 + \ldots + z_m^2$, where the $z_k$ are bilinear forms in the $X_i$ and $y_j$.[11] Then $S_{\mathbb{F}}(n) = S_{\mathbb{F}}(f_n)$ if $f_n$ denotes the left-hand side of (5.2).

Similarly, we define $\mathscr{B}_{\mathbb{F}}(f)$ as the minimal $m$ such that $f = z_1 z_1' + \ldots + z_m z_m'$, for some bilinear forms $z_k, z_k'$ in the $x_i$ and $y_j$. Over $\mathbb{F} = \mathbb{C}$, the quantities $\mathscr{B}_{\mathbb{F}}(f)$ and $S_{\mathbb{F}}(f)$ differ at most by a factor of two: $\mathscr{B}_{\mathbb{C}}(f) \leq S_{\mathbb{C}}(f) \leq 2\mathscr{B}_{\mathbb{C}}(f)$, which is easily seen.

Hrubes, Wigderson and Yehudayoff [HWY11] proved that a non-trivial lower bound on $\mathscr{B}_{\mathbb{F}}(f_n)$ for any explicit family $(f_n)$ of biquadratic polynomials over $\mathbb{F}$ implies an exponential lower bound on the size of noncommutative circuits for the noncommutative permanent, $\mathrm{per}_n := \sum_{\pi \in S_n} x_{1,\pi(1)} \cdots x_{n,\pi(n)}$.

**Theorem 5.1.** *Let $\mathbb{F}$ be a field with* $\mathrm{char}\,\mathbb{F} \neq 2$ *and* $(f_n)$ *be an explicit sequence of biquadratic polynomials over $\mathbb{F}$ such that $\mathscr{B}_{\mathbb{F}}(f_n) \geq \Omega(n^{1+\epsilon})$ for a constant $\epsilon > 0$. Then the noncommutative permanent* $\mathrm{per}_n$ *over $\mathbb{F}$ requires noncommutative circuits of size $2^{\Omega(n)}$.*

Explicit means here that there is an algorithm, which for given $n$ and exponent vector $\alpha$ computes the coefficient of $f_n$ at the monomial corresponding to $\alpha$, in time polynomial in $\log n$. For instance, the left-hand side of (5.2) defines an explicit family. So proving $S_{\mathbb{C}}(n) \geq \Omega(n^{1+\epsilon})$ would be sufficient. Interestingly, there are explicit families $(f_n)$ over $\mathbb{R}$ satisfying $S_{\mathbb{R}}(n) \geq \Omega(n^2)$, but unfortunately, their construction breaks down over $\mathbb{C}$, see [HWY11].

In view of the derandomization result achieved in the noncommutative model (Theorem 6.8 below), an unconditional version of Theorem 5.1 may be within reach.

### 5.5. **Constant depth circuits.**

As for Boolean circuits, it is natural to restrict to circuits of bounded depth. For this purpose, we modify Definition 2.1 by allowing gates of any indegree (unbounded fan-in). An addition gate $v$ with children $v_1, \ldots, v_a$ now computes the linear combination $\lambda_1 \hat{\Phi}_{v_1} + \ldots + \lambda_a \hat{\Phi}_{v_a}$ of the polynomials $\hat{\Phi}_{v_i}$ computed at $v_i$, where the field element $\lambda_i \in \mathbb{F}$ is though to be a label of the edge connecting $v_i$ with $v$. The size of such circuits is defined as the number of its edges.

The following observation is attributed to Ben-Or. Consider the identity $(t + x_1) \cdots (t + x_n) = t^n + \sum_{d=1}^{n} \mathrm{esym}_d(x) t^{n-d}$, where $\mathrm{esym}_d(x)$ denotes the $d$-th elementary symmetric polynomial. We evaluate this product at $n$ distinct values $t_1, \ldots, t_n \in \mathbb{F}$ and retrieve from this $\mathrm{esym}_1(x), \ldots, \mathrm{esym}_n(x)$ by interpolation. This shows that the elementary symmetric polynomial can be computed by a homogeneous depth-three formula of size $O(n^2)$ with the structure $\Sigma\Pi\Sigma$, in which the first layer has addition gates, the second layer (unbounded fan-in) product gates, and the third layer (unbounded fan-in) addition gates. This is striking, since $\mathrm{esym}_1 = x_1 + \ldots + x_n$ over $\mathbb{F}_2$ describes the parity function, for which it is known [Has86] that Boolean circuits of bounded depth require exponential size. The paper [SW01] proved for $\log n \leq d \leq 2n/3$ an $\Omega(n^2/d)$ lower bound for depth-three circuits, thus showing the optimality of Ben-Or's depth-three circuit. Recently, Chatterjee et al. [CKSV22] proved that arithmetic formulas of any depth computing $\mathrm{esym}_{n/10}(x)$ require size at least $\Omega(n^2)$, thus showing the optimality of Ben-Or's construction in a very strong sense.

---

[11]Biquadratic means that every monomial of $f$ has the form $x_{i_1} x_{i_2} y_{j_1} y_{j_2}$.

Remarkably, Grigoriev and Karpinski [GK99], using the method of partial derivatives, proved an exponential lower bound for depth-three circuits computing the determinant *over finite fields of constant size*. In fact, by the same technique, exponential lower bounds for elementary symmetric polynomials can be shown for depth-three circuits over small finite fields, see [Sap15, Thm. 10.2].

Agrawal and Vinay [AV08] showed that for answering the VP vs VNP question, it is enough to focus on circuits of depth four. Their result was first improved by Koiran [Koi12] and then by Tavenas [Tav15], who proved the following parallelization result in the spirit of [VSBR83] (compare Theorem 2.12).

**Theorem 5.2.** *Assume $f \in \mathbb{F}[X_1, \ldots, X_n]$ has degree $d$ and can be computed by an arithmetic circuit of size $s$. Then there is depth-four circuit of size $2^{O(\sqrt{d \log(ds) \log n})}$ computing $f$. This circuit can be assumed to have the structure $\Sigma\Pi\Pi^\alpha\Sigma\Pi\Pi^\beta$, where $\beta = O\big(\sqrt{d\frac{\log(ds)}{\log(n)}}\big)$ bounds the fan-in of the product gates at level one and $\alpha = O\big(\sqrt{d\frac{\log n}{\log(ds)}}\big)$ bounds the fan-in of the product gates at level three. Additionally, if $f$ is homogeneous, the circuit can be chosen to be homogenous.*

Therefore, if the $n \times n$ permanent can be computed by an arithmetic circuit of size polynomial in $n$, then it can also be computed by a $\Sigma\Pi\Pi^\alpha\Sigma\Pi\Pi^\alpha$ arithmetic circuit of size $2^{O(\sqrt{n}\log n)}$, where $\alpha = O(\sqrt{n})$. This appears to be close to optimal when comparing it with the following breakthrough lower bound result, obtained by Kayal et al. [GKKS14].

**Theorem 5.3.** *Any arithmetic circuit computing the $n \times n$ permanent (or the $n \times n$ determinant) over a field $\mathbb{F}$, having the structure $\Sigma\Pi\Pi^\alpha\Sigma\Pi\Pi^\alpha$ with $\alpha = O(\sqrt{n})$, necessarily has size at least $2^{\Omega(\sqrt{n})}$.*

Thus, as a consequence of Theorem 5.2, in order separate VP from VNP, it would be sufficient to raise the lower bound in the exponent from $\Omega(\sqrt{n})$ to $\omega(\sqrt{n}\log n)$. But note that the above theorem cannot distinguish between permanent and determinant. The paper [GKKS13] provides even a reduction to circuits of depth three with structure $\Sigma\Pi\Sigma$ (over $\mathbb{Q}$).

Behind the proof in [GKKS14] is a refinement of the rank of partial derivatives method, called *method of shifted partial derivatives*. In the language of algebraic geometry, one studies the growth of the Hilbert functions of Jacobian ideals. For the determinant, this is the ideal generated by subdeterminants of a certain size, which is well understood [CH94]. For the permanent, this amounts to the ideal of subpermanents, for which little is known. The limits of the method were explored in [ELSW18] who proved that the separation of VP and VNP cannot be achieved along this way.

After the publication of [GKKS14], the method of shifted partial derivatives created a flurry of activities and follow-up works, too many to be summarized here. For instance, it was shown [KLSS14, FLMS15, KS17] that the lower bounds in Theorem 5.3 also apply for iterated matrix multiplication, showing the optimality of Tavenas' depth reduction. We recall that the iterated matrix multiplication polynomial $\mathrm{IMM}_{n,d}$ is defined as the trace of the matrix product $A_1 \cdots A_d$, where the $A_i$ are $n \times n$ matrices whose entries are indeterminates, see Remark 2.8. Note that $\mathrm{IMM}_{n,d}$ is set-multilinear.

Very recently, Limaye et al. [LST22] managed to prove, for the first time, a superpolynomial lower bound against algebraic circuits of all constant depths, over all fields of characteristic 0. Below is a precise statement of this breakthrough result. Note that if $d_n = \Theta(\log n)$, the lower bound $n^{\Omega((\log n)^\varepsilon)}$ indeed grows superpolynomially in the number $N_n = \Theta(n^2 \log n)$ of input variables.

**Theorem 5.4.** *Let $d_n \leq 10^{-2} \log n$ and $\mathbb{F}$ be of characteristic zero. For each $\Delta \geq 1$ there is $\epsilon > 0$ such that every family of arithmetic circuits of depth at most $\Delta$, computing the iterated matrix multiplication polynomial $\mathrm{IMM}_{n,d_n}$ over $\mathbb{F}$, has size at least $n^{d_n^\epsilon}$.*

The sophisticated proof is based on an extension of "set multilinearization" [Raz13] and on the method of partial derivatives.

## 6. Derandomizing the polynomial identity testing problem

The idea of building pseudorandom generators from hard functions originated in cryptography and became a central topic of theoretical computer science. By a combinatorial construction, Nisan and Wigderson [NW94] showed how to construct pseudorandom generators from hard functions in EXPTIME. In an influential paper [KI04], Impagliazzo and Kabanets proved a converse: the elimination of randomness (briefly called *derandomization*) implies the existence of hard problem (see [AvM11] for a simplified proof). In these arguments, ideas from algebraic complexity theory enter in a central way. The following computational problem plays a key role.

**Definition 6.1.** The *arithmetic circuit identity testing problem (ACIT)* is the task to decide for a given constant-free arithmetic circuit whether it computes the zero polynomial in $\mathbb{Z}[X_1, \ldots, X_n]$.

It is well known [IM83] that ACIT can be solved in random polynomial time, more specifically, ACIT lies in coRP. The idea is to pick a sufficiently small random prime number $p$ and to evaluate the given arithmetic circuit at random values $x_1, \ldots, x_n \in \mathbb{F}_p$. The restriction of PIT to $n = 0$ (deciding whether an integer given by an arithmetic circuit equals zero) is sometimes called the EquSLP problem: it is easily seen to be polynomial time equivalent to ACIT, see [ABKPM09].

In the definition of ACIT, we allow circuits computing polynomials of exponentially high degree. To avoid difficulties as in discussion of VP versus $\mathrm{VP}_{\mathrm{nb}}$ in Section 4.2, we now exclude this by requiring the input circuit to be weakly-skew (or equivalently, to be an algebraic branching program), see Remark 2.22. We call the resulting problem the *polynomial identity testing problem (PIT)*.[12]

Impagliazzo and Kabanets [KI04] showed that the existence of a *deterministic* subexponential time algorithm for PIT implies the separation of complexity classes in the following way.

**Theorem 6.2.** *If* PIT *can be decided with a nondeterministic subexponential time algorithm, then there is a language in nondeterministic exponential time that cannot be decided by Boolean circuits of polynomial size* or *the permanent cannot be computed by arithmetic circuits of polynomial size.*

The question whether PIT has a deterministic polynomial time algorithm is considered one of the major open questions of complexity theory.

We consider now the following computational problem, which asks whether the determinant of a given matrix, whose entries are linear forms with integer coefficients, vanishes identically. (This problem was first studied by Edmonds [Edm67].)

**Definition 6.3.** The *symbolic determinant identity testing problem (SDIT)* is the task to decide for given integer matrices $A_1, \ldots, A_m \in \mathbb{Q}^{n \times n}$ whether the determinant of $x_1 A_1 + \ldots + x_m A_m$ vanishes identically. Here $x_1, \ldots, x_m$ denote commuting variables.

Valiant's proof for the completeness of determinant (see Section 2.5) implies that SDIT is polynomial time equivalent to PIT.

*Remark* 6.4. Over a fixed finite field $\mathbb{F}$, we may define SDIT as follows: for given $A_1, \ldots, A_m \in \mathbb{F}^{n \times n}$, decide whether $\det(x_1 A_1 + \ldots + x_m A_m) = 0$ for all $x \in \mathbb{F}^m$. This problem is coNP-complete, see [BFS99, Thm. 16]. (The proof follows easily from Valiant's reduction in the proof of Proposition 2.21.)

---

[12]In the literature, there does not seem to be a clear distinction between ACIT and PIT.

How could one possibly decide SDIT with an efficient deterministic algorithm? We denote by $\mathscr{C}(n,d,s) := \{f \in \mathbb{C}[x_1,\ldots,x_n] \mid L(f) \leq s\}$ the set of $n$-variate complex polynomials of degree at most $d$ and complexity at most $s$.

**Definition 6.5.** A polynomial map $G\colon \mathbb{Q}^k \to \mathbb{Q}^n$ is called *hitting set generator* for $\mathscr{C}(n,d,s)$ iff

$$\forall f \in \mathscr{C}(n,d,s) \setminus \{0\} \ \exists a \in \mathbb{Q}^k \quad f(G(a)) \neq 0.$$

The seed length of $G$ is defined as $k$ and the degree of $G$ is the maximum of the degrees of the components of $G$.

Heintz and Schnorr [HS82] proved the existence of hitting set generators for $\mathscr{C}(n,d,s)$ with small seed length and degree (certainly polynomially bounded in $n,d,s$). Consider the Zariski closure $\overline{\mathscr{C}}(n,d,s) := \{f \in \mathbb{C}[x_1,\ldots,x_n] \mid \underline{L}(f) \leq s\}$ of $\mathscr{C}(n,d,s)$. (Compare Definition 4.18 of border complexity.) The proof in [HS82] relies on bounding the dimension and degree of the affine variety $\overline{\mathscr{C}}(n,d,s)$, as done by Strassen in his pioneering work [Str74]. This even provides hitting set generators for $\overline{\mathscr{C}}(n,d,s)$.

The difficulty is to get explicit hitting set generators. Let us call a family $(G_{n,d,s})$ of hitting set generators for $\mathscr{C}(n,d,s)$ of constant seed length $k$ *explicit* if the map $(n,d,s) \mapsto G_{n,d,s}$ is computable in deterministic polynomial time. Here we assume that $n,d,s$ are encoded in unary, the degree of $G_{n,d,s}$ is polynomially bounded in $n,d,s$, and $G_{n,d,s}$ is given by the list of its rational coefficients.

We observe that the existence of such an explicit family of hitting set generators implies that the PIT problem can be solved in deterministic polynomial time. Indeed, for $f \in \mathscr{C}(n,d,s)$ given by an arithmetic circuit of size at most $s$ and formal degree at most $d$, we first compute $G_{n,d,s}$ and then evaluate the composition $f \circ G_{n,d,s}$ at the grid points in $\{0,1,\ldots,D\}^k$, where $D := d \deg G_{n,d,s}$. This is possible in polynomial time, since $k$ is assumed to be constant. Note $f = 0$ iff $f \circ G_{n,d,s} = 0$ since $G_{n,d,s}$ is a hitting set generator. By the Schwartz-Zippel lemma, the latter holds iff $f \circ G_{n,d,s}$ vanishes at all grid points,

Recently, the following remarkable result was obtained by Guo et al. [GKSS22]. It guarantees the existence of a hitting set generator assuming the existence of an explicit family of hard polynomials. We call a family $(P_d)$ of rational $k$-variate polynomials with $\deg P_d = d$ *explicit* iff the coefficient vector of $P_d$ can be computed from $d$ in polynomial time ($d$ encoded in unary). To simplify the statement we assume $n = d = s$.

**Theorem 6.6.** *Let $k$ and $\delta > 0$ be fixed and assume $(P_d)$ is an explicit family of $k$-variate rational polynomials such that $\deg P_d = d$ and $L(P_d) \geq d^\delta$ for all $d$. Then there exists an explicit family $G_{s,s,s}$ of hitting set generators for $\mathscr{C}(s,s,s)$. In particular, PIT can be solved in deterministic polynomial time.*

This considerably improves on the corresponding result in [KI04]. The proof is purely algebraic, while the construction in [KI04] was a mixture of a combinatorial construction (Nisan-Wigderson generator [NW94]) with Kaltofen's result on the complexity of factors of multivariate polynomials (see Theorem 3.2).

Recall that $\tau(f)$, for $f \in \mathbb{Z}[x]$, is defined as the minimal size of an arithmetic circuit computing $f$ from $x$ and the free constants $0, \pm 1$, while the smaller complexity measure $L(f)$ allows any constants for free. By the *strengthened $\tau$-conjecture* we understand (4.5) with $\tau(f)$ replaced by $L(f)$. An even bolder conjecture was suggested in [Bür01]. For a number field $\mathbb{F}$ and $f \in \mathbb{F}[x]$, let $N_d(f)$ denote the number of irreducible factors of $f$ in $\mathbb{F}[x]$ having degree at most $d$. The *L-conjecture* claims that there exist $c_{\mathbb{F}} > 0$ such that

$$(6.1) \qquad\qquad \forall f \in \mathbb{F}[x] \quad N_d(f) \leq (d + L(f))^{c_{\mathbb{F}}}.$$

In [Che02] it was observed that even a weaker version of the $L$-conjecture implies a quantitative version of the difficult torsion theorem for elliptic curves, which is stronger than what is known.

If the strengthened $\tau$-conjecture is true, then the family of Pochhammer-Wilkinson polynomials $P_s(x) := (x-1)\dots(x-s)$ satisfies $L(P_s) \geq s^\delta$ for some $\hat\delta > 0$. Applying Theorem 6.6 with $k = 1$ implies:

**Corollary 6.7.** *If the strengthened $\tau$-conjecture is true, then* PIT *has deterministic polynomial time algorithms.*

We now discuss the identity testing problem for computations in noncommutative polynomial rings $\mathbb{F}\langle x_1, \dots, x_n\rangle$. For noncommutative formulas, Raz and Shpilka [RS05] found a deterministic polynomial time algorithm (in the so-called white box model).

The paper [GGOW20] largely extends this, in particular also allowing divisions. While it is obvious how to adapt the definition of PIT to this model, it is a priori unclear what SDIT problem would mean in this setting, since the notion of the determinant is undefined. However, there is a "smallest" skew field containing $\mathbb{F}\langle x_1, \dots, x_n\rangle$ as a subring, which is unique up to isomorphism and called the *free skew field*, see [Coh85]. We define the noncommutative version of SDIT by requiring the square $n \times n$-matrix $\underline{A} := x_1 A_1 + \dots + x_m A_m$ to be invertible over the free skew field. The proof of equivalence of SDIT to PIT (basically Valiant's proof) carries over to the noncommutative setting. There are various equivalent characterizations of invertibility over the free skew-field, some of which are quite concrete, see [GGOW20, Thm. 1.4]. Over $\mathbb{F} = \mathbb{C}$, one of these characterizations says that the invertibility of $\underline{A}$ is equivalent to the matrix tuple $(A_1, \dots, A_m)$ lying in the *null-cone* of the left-right action of $\mathrm{SL}_n(\mathbb{C})$ on $m$-tuples of $n \times n$-matrices. Combining insights from computational invariant theory with an algorithm by Gurvits [Gur04], the following exciting result was proven in [GGOW20],

**Theorem 6.8.** *The noncommutative version of the* SDIT *over $\mathbb{Q}$ can be solved in deterministic polynomial time.*

Meanwhile, it was realized that this result comes out naturally from a theory of non-commutative optimization [BFG+19].

The connection between derandomization and proving complexity lower bounds has fuelled algebraic complexity theory. Many derandomization results have been obtained in restricted models of computations. For details, we refer the interested reader to the surveys by Shpilka and Yehudayoff [SY10] and Saptharishi [Sap15].

## 7. Geometric complexity theory

According to Corollary 2.33, the separation VBP $\neq$ VNP is equivalent to proving that the family of permanents is not a $p$-projection of the family of determinants. Following Mulmuley and Sohoni's influential work [MS01, MS08], we replace Valiant's combinatorial notion of projection (Definition 2.15) by a concept closer to algebra and geometry, retaining the symmetries of the determinant. This involves taking limits and requires coping with singularities: we always work over the field $\mathbb{C}$. This amounts to considering border complexity and attacking the stronger conjecture VNP $\not\subseteq \overline{\mathrm{VBP}}$.

At first glance, this might look like entering an even wilder and more complicated setting. However, at least implicitly, this is behind all known complexity lower bounds based on methods from algebraic geometry, see Strassen's pioneering work [Str74] and the comments at the beginning of Section 4.7.

We will see that VNP $\not\subseteq \overline{\mathrm{VBP}}$ can be equivalently rephrased as an orbit closure problem in the realm of geometric invariant theory with a high degree of symmetries. The hope is that the highly developed methods from invariant and representation theory provide new techniques to attack the elusive problem of separation of complexity classes.

### 7.1. Symmetries of the determinant.

The determinant family DET is VBP-complete according to Theorem 2.20. Are there any algebraic properties of the determinant that could be made responsible for it being easy to compute, but still universal in the class VBP?

The are various algebraic ways to characterize the determinant. It defines a group homomorphism $\det_n \colon \mathrm{GL}_n \to \mathbb{C}^*$. Moreover, it is well known that every group homomorphism $\mathrm{GL}_n \to \mathbb{C}^*$, given by a rational function is of the form $\det^k$ for some $k \in \mathbb{Z}$. In fact, Gaussian elimination makes essential use of the fact that $\det_n$ is a group homomorphism.

The following terminology from multilinear algebra is very useful when studing group actions. We denote by $\mathrm{Sym}^n V^*$ the vector space of degree $n$ homogeneous polynomial functions on a finite-dimensional complex vector space $V$. The group $\mathrm{GL}(V)$ acts on $\mathrm{Sym}^n V^*$ in the canonical way: $(g \cdot f)(v) := f(g^{-1} v)$ for $g \in G$, $f \in \mathrm{Sym}^n V^*$. The symmetries of $f \in \mathrm{Sym}^n V^*$ are characterized by its *stabilizer group* $\mathrm{stab}(f) := \{g \in G \mid g \cdot f = f\}$. Clearly, $\mu_n := \{z I \mid z^n = 1\}$ is contained in $\mathrm{stab}(f)$ and it is known that for generic $f$ equality holds, see [BI17, Thm. 2.3].

We assume now $V := \mathbb{C}^{n \times n}$, put $G := \mathrm{GL}(\mathbb{C}^{n^2})$, and view the determinant $\det_n$ as an element of $\mathrm{Sym}^n V^*$. It turns that $\det_n$ has a large stabilizer group, that we shall denote by $H$. Its elements are easy to guess: for $A, B \in \mathrm{SL}_n$ consider the following linear map given by matrix multiplication:

$$(7.1) \qquad\qquad g_{A,B} \colon \mathbb{C}^{n \times n} \to \mathbb{C}^{n \times n}, \ X \mapsto AXB.$$

Then $\det(AXB) = \det(A)\det(X)\det(B) = \det(X)$ by the homomorphism property of det, hence $g_{A,B} \in H$. The transposition $\tau \colon \mathbb{C}^{n \times n} \to \mathbb{C}^{n \times n}, \ X \mapsto X^T$ clearly also belongs to $H$. Frobenius [Fro97] proved that each element of $H$ is of the form $g_{A,B}$ or $\tau g_{A,B}$. (This result was rediscovered much later by Dieudonné [Die49].) The group $H$ is characterized by

$$(7.2) \qquad\qquad H \simeq (\mathrm{SL}_n \times \mathrm{SL}_n)/\mu_n \rtimes \mathbb{Z}_2.$$

This reveals that $H$ is a reductive group. It is a remarkable fact that, up to a scaling factor, $\det_n$ is uniquely determined by its stabilizer:

$$(7.3) \qquad\qquad H \subseteq \mathrm{stab}(f) \Longrightarrow \exists z \in \mathbb{C} \quad f = z \det_n.$$

This can be either verified directly or deduced from Theorem 7.6. So when replacing the polynomial $\det_n$ by its stabilizer $H$, no information gets lost.

*Remark* 7.1. Frobenius' characterization (7.2) of the symmetries of the determinant plays an essential role in the recent paper [MW21], which analyzes the symbolic determinant identity testing problem (Definition 6.3) from the point of view of geometric invariant theory. Makam and Wigderson proved that the variety

$$\mathrm{Sing}_{n,m} := \{(A_1, \ldots, A_m) \in (\mathbb{C}^{n \times n})^m \mid \forall x_1, \ldots, x_m \in \mathbb{C} \quad \det(x_1 A_1 + \ldots + x_m A_m) = 0\}$$

of singular matrix tuples is not a null cone for a reductive group action. This question arose since the corresponding variety of nonsingular tuples for noncommutative identity testing in fact is a null cone, which led to Theorem 6.8 as shown in [GGOW20].

## 7.2. Orbit closures.
We define the *orbit* and the *orbit closure* of $\det_n$ as

$$(7.4) \qquad\qquad \Omega_n^o := \mathrm{GL}_{n^2} \cdot \det_n \subseteq \mathrm{Sym}^n(\mathbb{C}^{n \times n})^*, \quad \Omega_n := \overline{\Omega_n^o} \subseteq \mathrm{Sym}^n(\mathbb{C}^{n \times n})^*.$$

The closure with respect to the Euclidean topology equals the closure with respect to the Zariski topology (same proof as for Theorem 4.19). Since $\Omega_n$ is closed under scalar multiplication, $\Omega_n$ is the affine cone over some projective variety, so we may equivalently think of $\Omega_n$ in terms of the corresponding complex projective variety. Note that

$$\dim \Omega_n = \dim \mathrm{GL}_{n^2} - \dim H = n^4 - 2n^2 + 2,$$

while $\dim \mathrm{Sym}^n \mathbb{C}^{n \times n} = \binom{n^2 + n - 1}{n}$. We may assume $n > 3$ since for $n = 2$ we have $\Omega_2 = \mathrm{Sym}^2(\mathbb{C}^{2 \times 2})^*$. Like any orbit, $\Omega_n^o$ is smooth. It is known that the *boundary* $\partial \Omega_n := \Omega_n \setminus \Omega_n^o$ is a hypersurface in $\Omega_n$. This follows by a general principle from the fact the stabilizer $H$ is reductive [BLMW11, §4.2]. But more is known [BI17, Prop. 3.10]: $\partial \Omega_n$ is the zero set in $\Omega_n$ with respect to the *fundamental* *invariant* of $\det_n$, which is the smallest nontrivial $\mathrm{SL}_{n^2}$-invariant of the coordinate ring $\mathbb{C}[\Omega_n]$, which is uniquely determined up to scaling. We also know [Kum13] that $\Omega_n$ is not a normal variety if $n > 2$; see [BI17, Cor. 3.18] for generalizations. So passing from the orbit $\Omega_n^o$ to the closure $\Omega_n$ presumably adds complicated singularities. For $n = 3$, the boundary of $\Omega_n$ has two irreducible components [HL16], but the boundary of $\Omega_4$ is already unknown, see [Hüt17].

The following insight from [MS01] is very important. In the language of geometric invariant theory, it expresses that $\det_n$ is polystable. The result follows readily with a refinement of the Hilbert-Mumford criterion [MFK94], see [BI17, Prop. 2.8].

**Theorem 7.2.** *The orbit of $\det_n$ under the action of $\mathrm{SL}_{n^2}$ is closed.*

Let us now explain why $\Omega_n$ naturally enters our setting. Suppose $\mathrm{per}_m(X) = \det(A(X))$, where the matrix $A(X)$ is of size $n$, with affine linear entries in $x_{11}, \ldots, x_{mm}$ and $m < n$. Homogenizing this equation with an additional variable $t$, we obtain

$$(7.5) \qquad t^{n-m}\mathrm{per}_m(X) = t^n \mathrm{per}_m(t^{-1}X) = t^n \det(A(t^{-1}X)) = \det(tA(t^{-1}X)).$$

The entries of the matrix $tA(t^{-1}X)$ are linear forms in $t$ and the $X$-variables. We call $t^{n-m}\mathrm{per}_m(X)$ the *padded permanent*.

The $n^2$ entries of $tA(t^{-1}X)$, arranged as a vector, may be thought of as being obtained by multiplying some matrix $L \in \mathbb{C}^{n^2 \times (m^2+1)}$ with $(x_{11}, \ldots, x_{mm}, t)^T$. Now think of $t$ as being one of the variables in $\{x_{11}, \ldots, x_{nn}\} \setminus \{x_{11}, \ldots, x_{mm}\}$. Then $L(x_{11}, \ldots, x_{mm}, t)^T = L'(x_{11}, \ldots, x_{nn})^T$, where $L'$ is obtained by appending $n^2 - m^2 - 1$ zero columns to $L$. We thus see that $t^{n-m}\mathrm{per}_m(X)$ is obtained from $\det_n$ by the substitution $L'$. Since $\mathrm{GL}_{n^2}$ is dense in $\mathbb{C}^{n \times n}$, we can approximate $L'$ arbitrarily closely by invertible matrices and hence we obtain

$$t^{n-m}\mathrm{per}_m(X) \in \Omega_n.$$

The following conjecture was stated by Mulmuley and Sohoni in [MS01].

**Conjecture 7.1.** *For all $c \in \mathbb{N}_{\geq 1}$ we have $t^{m^c - m}\mathrm{per}_m \notin \Omega_{m^c}$ for infinitely many $m$.*

The previous reasoning showed that if the determinantal complexity $\mathrm{dc}(m)$ is polynomially bounded in $m$, then Conjecture 7.1 is false. Therefore, by Valiant's completeness results, Conjecture 7.1 implies that $\mathsf{VNP} \not\subseteq \mathsf{VBP}$. By a closer look, one can show a more precise statement, see [BLMW11, Prop. 9.3.2].

**Theorem 7.3.** *Conjecture 7.1 is equivalent to the separation $\mathsf{VNP} \not\subseteq \underline{\mathsf{VBP}}$.*

**7.3. Explicit equations for $\mathsf{VP}$?** Suppose $f \in \mathbb{C}[x_{11}, \ldots, x_{mm}]$ is such that $t^{n-m}f$ does not lie in $\Omega_n$. How could one possibly prove this? Strassen [Str74] proposed the following general method (but not in the context of $\Omega_n$). Since $\Omega_n$ is a Zariski closed affine cone, there is a homogenous polynomial function $R \colon \mathrm{Sym}^n V^* \to \mathbb{C}$ vanishing on $\Omega_n$, but such that $R(f) \neq 0$. We know the dimension of $\Omega_n$ and can bound its degree via Bezout's theorem. With this information, one can infer the existence of a nonzero $R$ in the vanishing ideal $I_n$ of $\Omega_n$ of a not too large degree. (With a bit more work one can even make sure that $R$ has small integer coefficients.) With this method, the statement of Conjecture 7.1 can be proved when $\mathrm{per}_m$ is replaced by a polynomial

$$f = \sum_{\pi \in S_m} c(\pi) x_{1,\pi(1)} \cdots x_{m,\pi(m)},$$

whose coefficient vector $(c(\pi))$, e.g., is a permutation of $(\sqrt{1}, \sqrt{2}, \ldots, \sqrt{m!})$. This also works when the entries of $(c(\pi))$ are doubly exponentially fast growing integers. However, this technique is way too coarse for coping with a member $f$ of a $p$-definable family, see [Bür00a, §8.1].

The question of the existence of algebraic natural proof barriers [GKSS17, FSV18] is connected to the existence of efficiently computable equations $R$ for the complexity class $\mathsf{VP}$. In [CKR+20] it was shown that there are efficiently computable equations for polynomials in VP and VNP with small integer coefficients (and also over finite fields). On the other hand, the recent work [KRST22] shows that, without the assumption of bounded coefficients, VNP does not have equations in VP if the permanent is hard (in characteristic zero). The proof does not appear to extend to VP. Clearly, more work is required to clarify the situation!

7.4. **Representation theoretic obstructions.** Fortunately, there is additional structure in the setting of Section 7.2, coming from symmetries. The group $\mathrm{GL}(V)$ acts on the space of functions on $\mathrm{Sym}^n V^* \to \mathbb{C}$ and preserves the vanishing ideal $I_n$ of $\Omega_n$. While there is little hope to concretely know $I_n$, representation theory allows to classify the functions in $I_n$ according to discrete types. Specifically, we are interested in the number of linearly independent functions in $I_n$ of a fixed type.

Let us now describe this idea in more details. For this, we need to recall several facts from representation theory, see [FH91, GW09] for more information. Due to to lack of space, we have to be very brief.

Let $E$ denote a rational $\mathrm{GL}_m$-module. That is, $E$ is a finite-dimensional $\mathbb{C}$-vector space together with a linear action of $\mathrm{GL}_m$ given by a group homomorphism $\pi \colon \mathrm{GL}_m \to \mathrm{GL}(E)$, whose components are products of polynomial functions on $\mathbb{C}^{m \times m}$ with a power $\det(g)^k$ of the determinant ($k \in \mathbb{Z}$). Such $\pi$ is called a rational *representation* of $\mathrm{GL}_m$. The module $E$ is called *simple* (or $\pi$ is called *irreducible*) iff $E$ does not contain nontrivial invariant subspaces. Let $W$ be a simple $\mathrm{GL}_m$-module. The sum $E_W$ of the submodules of $E$ isomorphic to $W$ is called the $W$-*isotypical component* of $E$. Then $E_W$ is isomorphic to a direct sum of $m_E(W)$ many copies of $W$, denoted $E_W \simeq W^{\oplus m_E(W)}$ and $m_E(W)$ is called the *multiplicity* of $W$ in $E$. Moreover, a fundamental theorem states that $E$ splits as a direct sum of its isotypical components. (This expresses that the group $\mathrm{GL}_m$ is reductive.) So we have decompositions, running over the simple $W$,

$$E = \bigoplus_W E_W \simeq \bigoplus_W W^{\oplus m_E(W)}.$$

Suppose $\varphi \colon E \to F$ is a surjective module homomorphism, i.e., an equivariant linear map. Then the isotypical components are mapped onto isotypical components and the multiplicities can only drop:

$$(7.6) \qquad\qquad \varphi(E_W) = F_W, \quad m_F(W) \le m_E(W).$$

This observation, which is a simple consequence of Schur's lemma, is at the heart of the strategy of geometric complexity theory.

What we said so far applies to all representations of reductive groups. We now specialize to $G = \mathrm{GL}_\ell$. The isomorphism types of irreducible rational $\mathrm{GL}_\ell$-modules can be labelled by *highest weights*, which can be viewed as integer vectors $\lambda \in \mathbb{Z}^\ell$ such that $\lambda_1 \ge \cdots \ge \lambda_\ell$. The *Schur-Weyl module* $V_\lambda = V_\lambda(\mathrm{GL}_\ell)$ is a model for an irreducible $\mathrm{GL}_m$-module of highest weight $\lambda$ and can be explicitly constructed. We also call $\lambda$ the *type* of $V_\lambda$. For instance, $V_\lambda = \mathrm{Sym}^\delta \mathbb{C}^\ell$ if $\lambda = (\delta, 0, \ldots, 0)$ with $\delta \in \mathbb{N}$. At the other extreme, $V_\lambda = \mathbb{C}$ is one-dimensional with the operation $g \cdot 1 = \det(g)^\delta$ if $\lambda = (\delta, \ldots, \delta)$ for $\delta \in \mathbb{Z}$. Note that $\lambda$ is a partition if $\lambda_\ell \ge 0$. Its *length* is defined as $\ell(\lambda) := \#\{i \mid \lambda_i \ne 0\} \le \ell$ and $|\lambda| := \sum_i \lambda_i$ is called its *size*. We write $\lambda \vdash s$ to express that $\lambda$ is a partition of size $s$.

It is time to return to the setting of Section 7.1, where we viewed $\mathrm{Sym}^n V^*$ as the space of homogeneous degree $n$ polynomials on the complex vector space $V$. Thus $\det_n$ and $\mathrm{per}_n$ are elements of $\mathrm{Sym}^n V^*$, when $V = \mathbb{C}^{n \times n}$ and $\ell = n^2$. For the moment, assume more generally that $V = \mathbb{C}^\ell$. With a view toward equations as in Section 7.3, we now study homogeneous degree $d$ polynomial functions

$$\mathrm{Sym}^n V^* \to \mathbb{C},$$

which can be identified with the elements of the double symmetric power $\mathrm{Sym}^d \mathrm{Sym}^n V$. We use the canonical action of $\mathrm{GL}_\ell = \mathrm{GL}(V)$ on this space.

We now focus on the isotypical decomposition

$$(7.7) \qquad \mathrm{Sym}^d\mathrm{Sym}^n\mathbb{C}^\ell \simeq \bigoplus_{\lambda \vdash dn} V_\lambda^{\oplus \mathrm{pleth}_n(\lambda)}.$$

We are interested in the multiplicities $\mathrm{pleth}_n(\lambda) \in \mathbb{N}$, which are called *plethysm coefficients*. In the special case $\ell = 2$, this decomposition describes the invariants and covariants of binary forms of degree $n$, which was a subject of intense study in the 19th century: famous names like Cayley, Sylvester, Clebsch, Gordan, Hermite, Hilbert and others are associated with it, e.g., see [Sch68, KR84, Stu93]. Much less is known when $\ell$ is large. We refer to [FI20] for results on the complexity of computing plethysm coefficients. Among other things, it is shown there that deciding positivity of plethysm coefficients is NP-hard. This is interesting, due to the discovery that positivity of Littlewood-Richardson coefficients can be tested in polynomial time [DLM06, MNS12, BI13a], which led to the early GCT conjecture [Mul09] that this should also be possible for plethysm coefficients and Kronecker coefficients (see Section 7.5). For the latter this was refuted in [IMW17].

We now specialize to $V = \mathbb{C}^{n \times n}$ so that $\ell = n^2$. Recall from (7.4) the orbit closure $\Omega_n \subseteq \mathrm{Sym}^n V^*$ of $\det_n$. The homogenous coordinate ring $\mathbb{C}[\Omega_n] = \oplus_d \mathbb{C}[\Omega_n]_d$ is defined as the graded ring of polynomial functions on $\Omega_n$. It contains all the information on $\Omega_n$. Consider the surjective restriction homomorphism $\mathrm{Sym}\,\mathrm{Sym}^n\mathbb{C}^\ell \to \mathbb{C}[\Omega_n]$, whose kernel is the vanishing ideal $I_n$ of $\Omega_n$. Clearly, we have a $\mathrm{GL}_\ell$-action on $\mathbb{C}[\Omega_n]$ and the restriction homomorphism is equivariant. By our general reasonings, there is also a decomposition

$$(7.8) \qquad \mathbb{C}[\Omega_n]_d \simeq \bigoplus_{\lambda \vdash dn} K_n(\lambda)V_\lambda.$$

with multiplicities $K_n(\lambda)$, that we would like to call *gct-multiplicities*. Note $K_n(\lambda) \leq \mathrm{pleth}_n(\lambda)$.

After all these preparations we can now precisely describe the strategy. Suppose $h \in \mathrm{Sym}^n\mathbb{C}^\ell$ is a presumably hard function. For instance, $h$ may be a padded permanent $t^{n-m}\mathrm{per}_m$ (which only depends on $m^2 + 1$ variables). Consider the decomposition

$$\mathbb{C}[\overline{\mathrm{GL}_\ell \cdot h}] \simeq \bigoplus_{\lambda \vdash dn} V_\lambda^{\oplus \mathrm{mult}_h(\lambda)}$$

of the coordinate ring of the orbit closure of $h$. The assumption $h \in \Omega_n$ is equivalent to the inclusion of orbit closures

$$\overline{\mathrm{GL}_\ell \cdot h} \subseteq \Omega_n.$$

The restriction is a surjective equivariant ring homomorphism $\mathbb{C}[\Omega_n] \to \mathbb{C}[\overline{\mathrm{GL}_\ell \cdot h}]$, which via (7.6) implies the inequality of multiplicities

$$(7.9) \qquad \forall \lambda \vdash dn \quad \mathrm{mult}_h(\lambda) \leq K_n(\lambda).$$

A partition $\lambda$ violating this inequality is called *representation theoretic obstruction*.

The implementation of this strategy proved to be extremely difficult. The main obstacle is that hardly anything is known about the gct-multiplicities $K_n(\lambda)$. While the existence of suitable families of representation theoretic obstructions is sufficient for the separation VNP $\not\subseteq$ <u>VBP</u> of complexity classes, it is unclear whether representation theoretic obstruction are necessary. Any result on the principal feasibility of this strategy would be highly welcome!

We note that the outlined strategy of representation theoretic obstruction is completely general. Since $\Omega_n$ does not seem accessible at the moment, the strategy should be tested in simpler models of computation. For instance, one can study the Waring rank of forms (for which the partial derivatives method yields good exponential lower bounds). There, the analogue of $\Omega_n$ is the orbit closure of the power sum $X_1^n + \ldots + X_{n^2}^n$,

$$(7.10) \qquad \mathrm{PS}_n := \overline{\mathrm{GL}_{n^2} \cdot (X_1^n + \cdots + X_{n^2}^n)} \subseteq \mathrm{Sym}^n(\mathbb{C}^{n^2})^*.$$

which are also known as higher secant variety of the Veronese variety. There are some first positive results [DIP20, IK20] on the separating power of multiplicities in this setting.

### 7.5. Occurrence obstructions.

In [MS08] it was proposed to attack VNP $\not\subseteq$ $\underline{\text{VBP}}$ by showing a statement stronger than (7.9). Namely, to identify a highest weight $\lambda$ such that $K_n(\lambda) = 0$ and $\text{mult}_h(\lambda) > 0$, where $h = t^{n-m}\text{per}_m$ is the padded permanent. Such $\lambda$ are called *occurrence obstructions*. Note that $K_n(\lambda) = 0$ means that every function $R$ of type $\lambda$ in (7.7) vanishes on $\Omega_n$.

After a series of papers trying to implement this strategy, Bürgisser et al. [BIP19] finally showed that this is impossible.

**Theorem 7.4.** *Let* $n \geq m^{25}$ *and* $h := t^{n-m}\text{per}_m$. *If* $\lambda \vdash nd$ *satisfies* $\text{mult}_h(\lambda) > 0$, *then* $K_n(\lambda) > 0$.

Despite this negative result, it is worthwhile to explain some of the insights obtained along this way, since they certainly are relevant for progressing further (and are beautiful mathematics).

An important ingredient of the proof is an insight by Landsberg and Kadish [KL14], who showed that for any $h$ of the form $h = t^{n-m}f$, the condition $\text{mult}_h(\lambda) > 0$ enforces that $\lambda$ has relatively few rows and almost all of its boxes are in its first row.

**Proposition 7.5.** *Let* $f$ *be a form of degree* $m$ *in* $\ell \leq n^2$ *variables, and* $h := t^{n-m}f$. *If* $\text{mult}_h(\lambda) > 0$, *then* $\ell(\lambda) \leq \ell + 1$ *and* $\lambda_1 \geq |\lambda|(1 - m/n)$.

The first attempts towards analyzing occurrence obstructions relied on one further, but quite drastic simplification: to replace $\mathbb{C}[\Omega_n]$ by the coordinate ring $\mathbb{C}[\Omega_n^o]$ of the *orbit*, for the purpose of analysis of multiplicities.

The coordinate ring $\mathbb{C}[\Omega_n^o]$ of the orbit contains $\mathbb{C}[\Omega]$ as a subalgebra: one can think of the elements in $\mathbb{C}[\Omega_n]$ as those regular functions on $\Omega_n^o$ that can be extended to a regular function on $\Omega_n$. It should be noted that this is a stronger requirement than just a continuous extension; the reason behind is that the variety $\Omega_n$ is not normal, see [BI17, Prop. 3.10]. Consider the decomposition of $\mathbb{C}[\Omega_n]_d$ into irreducible $\text{GL}_{n^2}$-modules:

$$(7.11) \qquad\qquad \mathbb{C}[\Omega_n^o]_d \simeq \bigoplus_{|\lambda|=dn} k_n(\lambda)V_\lambda.$$

Obviously, $K_n(\lambda) \leq k_n(\lambda)$. The Peter-Weyl Theorem is a fundamental result of harmonic analysis. Its algebraic version allows to express the multiplicities $k_n(\lambda)$ as the dimension of the space of $H$-invariants in $V_\lambda$:

$$(7.12) \qquad\qquad k_n(\lambda) = \dim V_\lambda^H.$$

It turns out that the multiplicities $k_n(\lambda)$ are quantities familiar to researchers in algebraic combinatorics. Their description involves representations of the *symmetric group* $S_N$. To each partition $\lambda$ of size $|\lambda| = N$, there corresponds an irreducible $S_N$-module, called *Specht module*. Let $\lambda, \mu, \nu$ be partition of $N$. The corresponding *Kronecker coefficient* is defined as the dimension of the space of invariants of the tensor product $[\lambda] \otimes [\mu] \otimes [\nu]$:

$$k(\lambda, \mu, \nu) = \dim \left([\lambda] \otimes [\mu] \otimes [\nu]\right)^{S_N}.$$

Unfortunately, despite being fundamental, Kronecker coefficients are not well understood. We focus on the special case where $\mu$ and $\nu$ equal the rectangular partition $n \times d := (d, \ldots, d)$ of size $dn$.

The following result was essentially shown in [MS08], see [BLMW11, Prop. 5.2.1] for the equality statement. The proof is based on (7.12), uses the knowledge of $H$ from (7.2), and applies Schur-Weyl duality.

**Theorem 7.6.** *We have* $k_n(\lambda) \leq k(\lambda, n \times d, n \times d)$. *More precisely,* $k_n(\lambda)$ *equals the dimension of the space of* $S_{nd}$-*invariants in the tensor product* $[\lambda] \otimes [n \times d] \otimes [n \times d]$ *of* $S_{dn}$-*modules, that are also invariant under swapping the second and third factors.*

The discovery of the connection to Kronecker coefficients has stimulated research in algebraic combinatorics on these quantities. Since this is outside the scope of this overview, let us just refer to [Bür16] and the references given there.

We return to the plan of proving VNP $\not\subseteq$ $\underline{\text{VBP}}$ via occurrence obstructions in the simplified form, where one tries to exhibit families of highest weight $\lambda$ such that $k_n(\lambda) = 0$ and $\text{mult}_h(\lambda) > 0$. One may call such $\lambda$ *orbit occurrence obstructions*. (Recall that $k_n(\lambda) = 0$ implies $K_n(\lambda) = 0$.) Ikenmeyer and Panova [IP17] proved that that this is impossible. More specifically, they proved that if $n > 3m^4$ and $\text{mult}_h(\lambda) > 0$ for any $h$ of the form $h = t^{n-m}f$, then $k_n(\lambda) > 0$.

Soon after [IP17], Theorem 7.4 was proved in [BIP19]. While its proof was greatly inspired by [IP17], it differs vastly in the details and techniques. In particular, the link to Kronecker coefficients is not used in the proof. Remarkably, the only information used about the orbit closure $\Omega_n$ in the proof is that it contains certain padded power sums. In addition, as in [IP17], the only information used about the orbit closure of the padded permanent $t^{n-m}\text{per}_m$ is that it has the shape $t^{n-m}f$ for some $f \in \text{Sym}^m V^*$ (this enters via Proposition 7.5). So [BIP19] obtained the dramatic result that the strategy of occurrence obstructions cannot even be used to prove exponential lower bounds for the Waring rank against padded polynomials!

The proofs of the impossibility results mentioned so far heavily rely on the drastic effect the padding has on the multiplicities (Proposition 7.5). This can be avoided when working with the VBP-complete iterated matrix multiplication family $\text{IMM}_{n,m}$, depending on the two parameters size $n$ and degree $m$, instead of the determinant family (see Corollary 2.24). Let $\text{IMM}_{n,m} = \text{trace}(X_1 \cdots X_m)$, where the $X_i$ are $n \times n$-matrices of independent variables. The goal is to show that the smallest $n$, for which $\text{per}_m$ lies in the orbit closure of $\text{IMM}_{n,m}$, grows superpolynomially in $m$. It is unknown whether the method of occurrence obstructions can achieve this. We refer to [DGI+24b] for details on this homogeneous setting.

The paper [GIP17] studies the analogous question for the VBP-complete matrix powering family $\text{trace}(X^m)$. It is shown that with *orbit* occurrence obstructions, not even superlinear lower bounds can be shown! While this makes it unlikely that occurrence obstructions could do the job, this is not excluded by this paper.

Despite these negative results, it should be mentioned that in [BI11, BI13b], occurence obstructions were analyzed for proving the lower bound $\frac{3}{2}m^2 - 2$ on the border rank of $m \times m$ matrix multiplication. (In fact, this works with orbit occurence obstructions.) The best known lower bound is $2m^2 - \log_2 m - 1$, as shown in [LMe18], improving on [LO15].

### 7.6. Summary and outlook.

This overview should give plain evidence of the enormous impact of Les Valiant's algebraic reformulation of the P versus NP problem on the development of algebraic complexity theory. The area is flourishing and the wide range of connections established to different areas, in particular pure mathematics, is impressive.

Institute of Mathematics, Technische Universität Berlin
*Email address*: pbuerg@math.tu-berlin.de